\documentclass[twocolumn]{aastex7}

\makeatletter
\renewcommand{\frontmatter@title@above}{}
\makeatother

\usepackage{threeparttable} 
\usepackage{graphicx}      
\usepackage{xcolor}   
\usepackage[dvipsnames]{xcolor}
\usepackage{xspace}
\usepackage{subfigure}
\usepackage{booktabs}
\usepackage{subcaption}
\usepackage{hyperref}
\usepackage{wrapfig}
\usepackage{sidecap}
\usepackage{float}
\usepackage{placeins}
\usepackage{capt-of} 
\usepackage{makecell}

\usepackage{amssymb, amsmath, mathtools}
\usepackage{natbib}
\usepackage{enumerate}

\usepackage{latexsym}
\usepackage[version=4]{mhchem} 
\usepackage{xspace}

\newcommand{\fleck}{\texttt{fleck}\xspace}
\newcommand{\spotrod}{\texttt{spotrod}\xspace}

\newcommand\degree{\degr}
\newcommand\degrees\degree

\newcommand\eureka{\texttt{Eureka!}\xspace}

\newcommand{\poseidon}{\texttt{POSEIDON}\xspace}

\DeclareSymbolFont{UPM}{U}{eur}{m}{n}
\DeclareMathSymbol{\umu}{0}{UPM}{"16}
\let\oldumu=\umu
\renewcommand\umu{\ifmmode\oldumu\else\math{\oldumu}\fi}

\newcommand\microns \micron

\let\oldsim=\sim
\renewcommand\sim{\ifmmode\oldsim\else\math{\oldsim}\fi}
\let\oldpm=\pm
\renewcommand\pm{\ifmmode\oldpm\else\math{\oldpm}\fi}
\newcommand\by{\ifmmode\times\else\math{\times}\fi}

\newbox{\wdbox}
\renewcommand\c{\setbox\wdbox=\hbox{,}\hspace{\wd\wdbox}}
\renewcommand\i{\setbox\wdbox=\hbox{i}\hspace{\wd\wdbox}}
\newcommand\n{\hspace{0.5em}}

\newcount\timect
\newcount\hourct
\newcount\minct
\newcommand\now{\timect=\time \divide\timect by 60
         \hourct=\timect \multiply\hourct by 60
         \minct=\time \advance\minct by -\hourct
         \number\timect:\ifnum \minct < 10 0\fi\number\minct}

\catcode`@=11

\newcommand\comment[1]{}
\newcommand\commenton{\catcode`\%=14}
\newcommand\commentoff{\catcode`\%=12}

\renewcommand\math[1]{$#1$}
\newcommand\mathshifton{\catcode`\$=3}
\newcommand\mathshiftoff{\catcode`\$=12}

\comment{the backslash is necessary}

\comment{alignment tab}

\let\atab=&
\newcommand\atabon{\catcode`\&=4}
\newcommand\ataboff{\catcode`\&=12}

\let\oldmsp=\sp
\let\oldmsb=\sb
\def\sp#1{\ifmmode
           \oldmsp{#1}%
         \else\strut\raise.85ex\hbox{\scriptsize #1}\fi}
\def\sb#1{\ifmmode
           \oldmsb{#1}%
         \else\strut\raise-.54ex\hbox{\scriptsize #1}\fi}
\newbox\@sp
\newbox\@sb
\def\sbp#1#2{\ifmmode%
           \oldmsb{#1}\oldmsp{#2}%
         \else
           \setbox\@sb=\hbox{\sb{#1}}%
           \setbox\@sp=\hbox{\sp{#2}}%
           \rlap{\copy\@sb}\copy\@sp
           \ifdim \wd\@sb >\wd\@sp
             \hskip -\wd\@sp \hskip \wd\@sb
           \fi
        \fi}
\def\msp#1{\ifmmode
           \oldmsp{#1}
         \else \math{\oldmsp{#1}}\fi}
\def\msb#1{\ifmmode
           \oldmsb{#1}
         \else \math{\oldmsb{#1}}\fi}
\def\supon{\catcode`\^=7}
\def\supoff{\catcode`\^=12}
\def\subon{\catcode`\_=8}
\def\suboff{\catcode`\_=12}
\def\supsubon{\supon \subon}
\def\supsuboff{\supoff \suboff}

\newcommand\actcharon{\catcode`\~=13}
\newcommand\actcharoff{\catcode`\~=12}

\newcommand\paramon{\catcode`\#=6}
\newcommand\paramoff{\catcode`\#=12}

\comment{And now to turn us totally on and off...}

\newcommand\reservedcharson{\commenton \mathshifton \atabon \supsubon \actcharon
	\paramon}

\newcommand\reservedcharsoff{\commentoff \mathshiftoff \ataboff
	\supsuboff \actcharoff \paramoff}

\catcode`@=12
\reservedcharsoff

\reservedcharson

\comment{ Must have ONLY ONE of these... trust these macros, they work

}

\newcommand{\squishlist}{
 \begin{list}{$\bullet$}
  { \setlength{\itemsep}{0pt}
     \setlength{\parsep}{0pt}
     \setlength{\topsep}{0pt}
     \setlength{\partopsep}{0pt}
     \setlength{\leftmargin}{2.0em}
     \setlength{\labelwidth}{1.5em}
     \setlength{\labelsep}{0.5em} } }

\newcommand{\squishlisttwo}{
 \begin{list}{$\bullet$}
  { \setlength{\itemsep}{1pt}
     \setlength{\parsep}{3pt}
     \setlength{\topsep}{3pt}
     \setlength{\partopsep}{0pt}
     \setlength{\leftmargin}{2.0em}
     \setlength{\labelwidth}{1.5em}
     \setlength{\labelsep}{0.5em} } }

\newcommand{\squishend}{
  \end{list}  }

\reservedcharson
\actcharon
\graphicspath{{./}{Figures/}}

\newcommand{\planetname}{HATS-75~b}

\begin{document}

\onecolumngrid

\begin{center}
    \large\textbf{GEMS JWST: HATS-75 b \\ 
    A giant planet with a sub-solar metallicity atmosphere orbiting an M-dwarf}\\
\end{center}

\begin{center}
    Reza Ashtari$^{1}$, Jacob Lustig-Yaeger$^{1}$, Jessica Libby-Roberts$^{2,3,4}$, Simon M\"uller$^{5}$, \\
    Shubham Kanodia$^{6}$, Kevin B. Stevenson$^{1}$, Caleb I.\ Ca\~nas$^{7}$, Giannina Guzm\'an Caloca$^{8,7}$, \\
    Nicole L.\ Wallack$^{6}$, Megan Delamer$^{2,3}$, Anjali A.\ A.\ Piette$^{9}$, Suvrath Mahadevan$^{2,3}$, \\
    Ian Czekala$^{10}$, Te Han$^{11}$, Ravit Helled$^{5}$
\end{center}

\begin{center}
\textit{
$^{1}$Johns Hopkins University Applied Physics Laboratory, Laurel, MD, USA\\
$^{2}$Department of Astronomy \& Astrophysics, The Pennsylvania State University, University Park, PA, USA\\
$^{3}$Center for Exoplanets and Habitable Worlds, The Pennsylvania State University, University Park, PA, USA\\
$^{4}$Department of Physics \& Astronomy, University of Tampa, Tampa, FL, USA\\
$^{5}$Department of Astrophysics, University of Z\"urich, Z\"urich, Switzerland\\
$^{6}$Earth and Planets Laboratory, Carnegie Science, Washington, DC, USA\\
$^{7}$NASA Goddard Space Flight Center, Greenbelt, MD, USA\\
$^{8}$Department of Astronomy, University of Maryland, College Park, MD, USA\\
$^{9}$School of Physics \& Astronomy, University of Birmingham, Birmingham, UK\\
$^{10}$School of Physics \& Astronomy, University of St.\ Andrews, St.\ Andrews, UK\\
$^{11}$Department of Physics \& Astronomy, University of California, Irvine, CA, USA
}
\end{center}

\vspace{0.8cm}
\begin{center}
    \textit{Accepted to AJ}
\end{center}

\begin{abstract}
\planetname{} is one of the recently discovered Giant Exoplanets orbiting M-dwarf Stars (GEMS) with a transmission spectrum shaped by both its atmosphere and the active stellar surface it transits. As part of a JWST program studying 7 GEMS, we observed three transits of \planetname{} with the NIRSpec PRISM instrument (0.6–5.3~$\mu$m). The planet's spectra exhibit a slightly larger transit depth at shorter wavelengths, indicative of hazes or stellar contamination due to stellar heterogeneities outside the transit chord, i.e., the transit light source (TLS) effect. While both a hazy atmospheric model or TLS model can replicate the transmission spectrum, independent evidence (.e.g, stellar rotation, spot-crossing events) favors a model that includes contamination from unocculted starspots and faculae. Within this stellar heterogeneity / TLS-based framework, atmospheric retrievals yield remarkably low atmospheric metallicity ($\rm log~[M/H]=-1.74^{+0.92}_{-0.76}$) and super-solar carbon-to-oxygen ($\rm C/O=1.04^{+0.40}_{-0.09}$), which paired with a best-fit interior model with bulk metallicity of Z$_p$ = 0.20 $\pm$ 0.04, implies poor vertical mixing within the planet. Retrievals also detect robust absorption signatures of CH$_4$, CO, and CO$_2$. We obtain only an upper limit for H$_2$O, consistent with its atmospheric spectral features being masked by stellar contamination.  These results underscore the importance of accounting for stellar heterogeneity when interpreting exoplanet transmission spectra and highlight \planetname{} as a significant asset to our understanding of giant exoplanets around M-dwarfs with JWST.
\end{abstract}

\keywords{M-dwarf stars: Exoplanet atmospheres: Extrasolar gaseous giant planets: Exoplanet atmospheric evolution: Planet formation\\ \\ \\ \\ \\ \\ \newline}

\section{Introduction}
\label{sec:intro}

M-dwarf stellar systems are opportune testing grounds for planet formation theories. These low-mass stars, which span roughly $0.08~M_\odot$ to $0.6~M_\odot$ with temperatures between $\sim2600~$K and $\sim4000~$K \citep{PecautMamajek2013}, are expected, on average, to host proportionally lower-mass protoplanetary disks \citep{Andrews2013,Pascucci2016, Manara2023}. Under the canonical core–accretion paradigm \citep{Mizuno1980,Pollack1996}, assembling the $\sim10~M_\oplus$ solid core necessary to trigger runaway gaseous accretion becomes challenging in such disks because the orbital timescales are longer and the disk mass is reduced compared to solar-type hosts \citep{Laughlin2004,IdaLin2005}. This means that by the time a sufficiently massive core could form, the gas reservoir may already be depleted. An alternative route via gravitational instability has been proposed \citep[e.g.,][]{Boss1997,Boss2006, boss_forming_2023}, but this mechanism typically requires a very massive, cool disk early in the protostellar phase, conditions that are currently not empirically well constrained for M-dwarf systems. These theoretical hurdles help explain why giant planets around M-dwarfs have long been regarded as rare \citep{burn_new_2021}.

Observational surveys corroborate the scarcity of giant planets orbiting M-dwarf stars. Radial velocity campaigns have discovered only a handful of gas giants around M-dwarfs \citep[e.g.,][]{Endl2006,Sabotta2021}, implying occurrence rates well below those for FGK hosts. Until recently, transit surveys provided few constraints because M-dwarfs comprised a small fraction of the target lists \citep{kanodia_searching_2024}. However, NASA’s TESS mission and ground-based surveys such as HATSouth and MANGOS have begun to unearth a modest but growing sample of short-period “Giant Exoplanets around M-dwarf Stars” \citep[GEMS;][]{Jordan2022, Glusman2025, mangos2025,  kanodia_searching_2024}. Early estimates place the occurrence of hot Jupiter analogues (radius $>8~R_\oplus$, period $<10$~days) around M-dwarfs at only $\sim0.1$–-$0.3\%$ \citep{Glusman2025}, yet each detection is invaluable for testing planet formation models in this mass-starved regime \citep{kanodia_toi-5205b_2023}.

Although only $\sim$30 transiting GEMS are currently confirmed, these rare planets provide critical leverage for comparing giant planet demographics across the stellar mass spectrum \citep{kanodia_transiting_2024}. Several recently identified systems orbit mid-to-late M-dwarfs ($M_\star \lesssim0.4 M_\odot$), where core–accretion theory would predict near-zero giant yield for typical disk masses unless nonstandard mechanisms, such as early disk fragmentation, intervene \citep{burn_new_2021}. By measuring masses and radii for GEMS, we can examine whether their bulk properties (e.g., densities, core mass fractions) systematically differ from hot Jupiters around FGK hosts \citep{kanodia_transiting_2024}. This growing sample will allow us to test whether GEMS exhibit distinct structural or atmospheric properties compared to their FGK counterpart \citep{Canas2025}.

Pursuing this motivation, we present three transit observations of \planetname{} using NIRSpec/PRISM as part of the GEMS Cycle 2 program (GO 3171) ---   \textit{Red Dwarfs and the Seven Giants: First Insights Into the Atmospheres of Giant Exoplanets around M-dwarf Stars} --- a large survey targeting short-period giant planets orbiting M-dwarfs in transmission \citep{kanodia_red_2023,Canas2025}. Discovered by the HATSouth network and TESS in 2021 \citep{Jordan2022}, \planetname{} is a gas giant with a mass of $\approx0.49~M_{\rm J}$ orbiting a $M_\star\approx0.6~M_\odot$, $T_{\rm eff}\approx3790~$ M dwarf on a 2.78-day period. 

Our observations offer an opportunity not only to characterize the atmospheric composition of \planetname{}, but also to explore how stellar heterogeneity influences the interpretation of transmission spectra in low-mass star systems. Section \ref{sec:data} describes our observational program and the JWST transit observations, along with the data reduction. In Section \ref{sec:retrievals}, we outline our analysis of the resulting spectra, employing a Bayesian retrieval framework to infer and constrain \planetname{}’s atmospheric properties. We then leverage the retrieved atmospheric model to gain new insight into the planet’s composition. Section \ref{sec:bulk} presents constraints of the planet’s atmospheric metallicity and uses interior structure modeling, estimating \planetname{}’s age, interior temperature, and bulk density. Section \ref{sec:discuss} discusses the implications of these findings, placing \planetname{} in the broader context of giant planet formation around M-dwarfs and highlighting how this benchmark system can refine our understanding of planet formation in low-mass stellar environments.

\section{Data} 
\label{sec:data}

\subsection{JWST Observations}
\label{sec:data:jwst}

Transit observations of \planetname{} presented here were made using the James Webb Space Telescope's (JWST's) \citep{Gardner2006} NIRSpec/PRISM Bright Object Time Series (BOTS) mode \citep{Jakobsen2022} as part of the larger Giant Exoplanets around M-Dwarf Stars JWST Cycle 2 survey\footnote{GO 3171; PI: Kanodia, Ca\~nas, Libby-Roberts. \\ GEMS team website: \href{https://gemsjwst.github.io/index.html}{gemsjwst.github.io}}. We observed three transits in total of \planetname{} on 2024 August 9 (Visit 1), 2024 August 15 (Visit 2), and 2024 August 17 (Visit 3) UT. Each observation utilized the NIRSpec/PRISM SUB512 subarray with 5 groups per integration, 14402 integrations per exposure yielding a total exposure time of $\sim5.5$ hours per visit. All pixel flux-levels remained less than the 80\% full-well depth.While this lies slightly above the $70-75\%$ ''safe regime'' for non-linearity designated by \citealt{Carter2024}, we found no evidence of non-linearity affecting the data.

With a J-magnitude of 12.5, we used HATS-75 for target acquisition. Every transit observation consisted of a pre-transit baseline of $\sim$ 2.5 hours which allowed for instrument settling followed by the 2 hour transit and an additional $\sim$1.5 hour post-transit baseline.

\subsection{Data Reduction and Light Curve Analysis}
\label{sec:data:reduction}

We performed two independent reductions of each HATS-75 observation to ensure repeatability. The first reduction, \textit{Optimized Eureka! $+$ fleck}, uses an automated pipeline for data reduction using \eureka \citep{Bell2022, Ashtari2025}. Transits and star spots apparent in the transits were then modeled using the \texttt{batman} \cite{Kreidberg2015} and \texttt{fleck} \citep{Morris2022} packages and the posteriors sampled with the \texttt{emcee} sampler \citep{emcee}. The second data reduction, \textit{Manual Eureka! $+$ spotrod}, also uses \eureka \citep{Bell2022}, but with reductions optimized manually. Transits from this method are then fitted using the \texttt{spotrod} star spot model \citep{Beky2014} with posteriors sampled using the \texttt{dynesty} dynamic nested sampler \citep{Speagle2020}.

\subsubsection{Optimized Eureka! $+$ fleck}
\label{subsec:data:reduction1} 

\begin{figure*}[t!]
  \centering
  \includegraphics[width=1.0\textwidth,trim={0.6cm 0.6cm 0.6cm 0.6cm},clip]{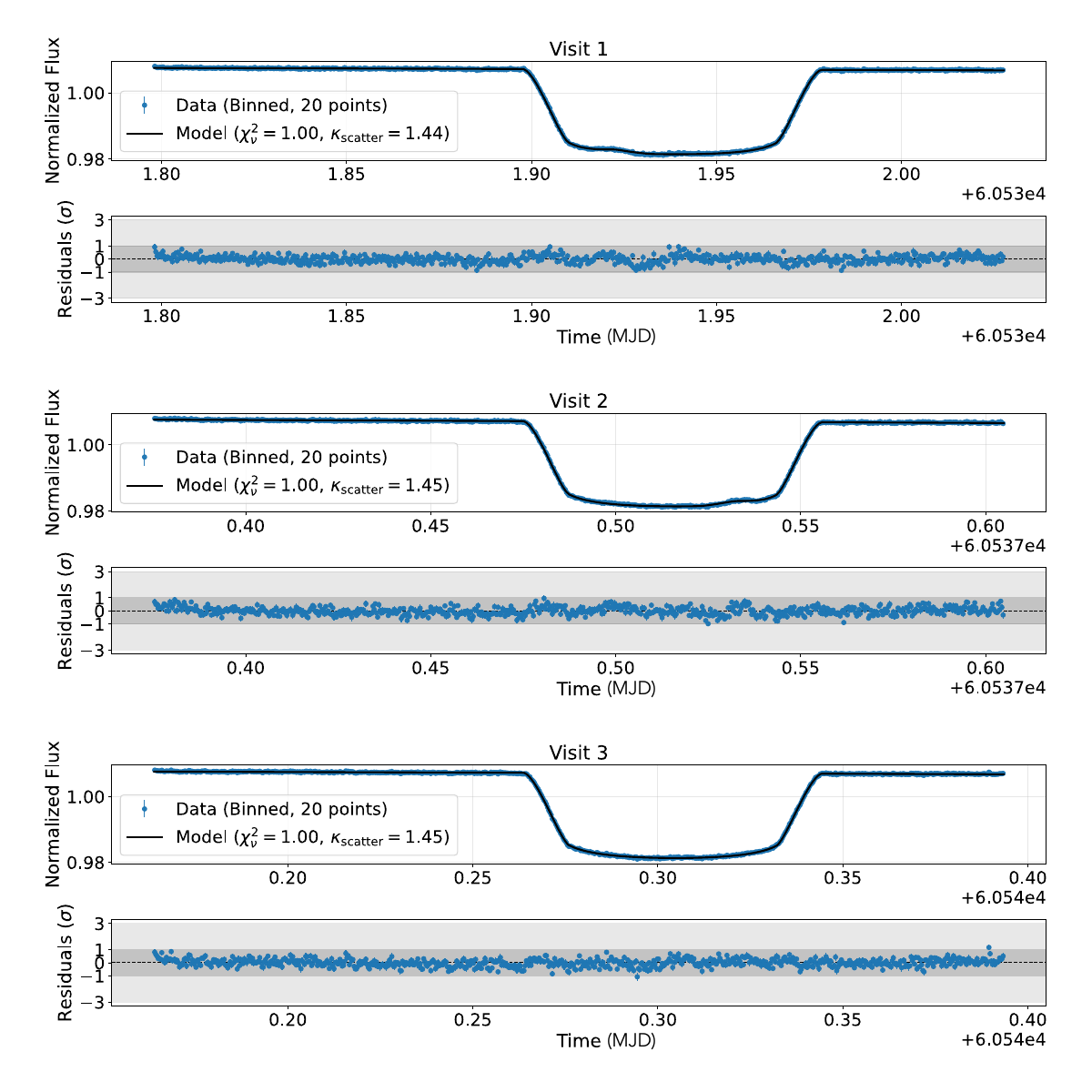}
  \caption{White light curves and best-fit transit models for Visits 1 (top), 2 (middle), and 3 (bottom) of the optimized \eureka\texttt{+ }\fleck dataset. Residuals for the model-fits are provided below the light curve of the same transit to show time-correlated noise. The bump near the ingress of Visit 1, to the left of the transit center ($\sim1.92 + 6.053e4$ MJD), represents the starspot illustrated in the top plot of \autoref{fig:spots_fleck}. Similarly, the raised feature near the egress of Visit 2, to the right of the transit center ($\sim0.54 + 6.0537e4$ MJD), represents the starspot illustrated in the bottom subfigure of \autoref{fig:spots_fleck}. $\chi^2_{\nu}$ and scatter multiplier values --- $\kappa_{scatter}$ --- \citep{Bell2022, Morris2022} are provided as metrics for the model-fit and necessary error inflation (to planet radius estimates) during light curve fitting.}
  \label{fig:wlc_visits}
\end{figure*}

Data reductions for all three transits were performed using a prototype data optimizer for \eureka v1.1.2 \citep{Ashtari2025} combined with the JWST calibration pipeline, \texttt{jwst} v1.18.0 (CRDS pmap v1364) \citep{jwst.calibration}. Stages 1, 3 and 4 are parametrically optimized using the optimizer described in \cite{Ashtari2025}. Stage 2 is executed using the parameter values from the Eureka! control file (\texttt{ecf}). Stages 1, 3 and 4 have several parameters in the \texttt{ecf}'s that define each step's execution in ramp fitting, background subtraction, spectral extraction, and outlier rejection. For each parameter, the optimizer sequentially sweeps through a specified range of values, and calculates a fitness score: the median absolute difference (MAD) of the spectroscopic light curves generated. The fitness score drives the optimizer to choose the best \texttt{.ecf} parameters producing time-series data with minimized scatter and reduced outliers (see \cite{Ashtari2025} for more detail). The parameter producing the lowest fitness score is selected as the best value, before continuing to optimize the next \texttt{.ecf} parameter. This process of sequential, parametric optimization is shown in \autoref{fig:optimization}. To accelerate this process, iterative optimization runs are performed on the last segment of Visit 3, as the transit appeared to be least affected by star spot crossings relative to Visits 1 and 2 (evident in \autoref{fig:wlc_visits}). Once all parameters are optimized, all stages are rerun on all segments of Visit 3. Given the same source, observation and instrument configurations, these same optimized values (shared in \autoref{fig:optimization} are then used to generate the light curves for Visits 1 and 2.

We fit the white light curves using \texttt{Eureka!} in combination with the \texttt{batman} transit and \texttt{fleck} star spot model using the \texttt{emcee} sampler. Eccentricity, and argument of periastron are fixed to values from \citet{Jordan2022}, while the planet-to-star radius ratio ($R_p/R_s$) and mid-transit time ($t_0$) are treated as free parameters with uniform priors. Due to computational limitations of jointly fitting the light curves of multiple transits while simulataneously modeling starspots, orbital period remains fixed to the value from \citet{Jordan2022}. The inclination ($i$) and scaled semi-major axis ($a/R_s$) are free with Gaussian priors informed by the discovery analysis of \citet{Jordan2022}. Quadratic limb darkening coefficients ($u_1$, $u_2$) are also allowed to vary under uniform priors. Star spots are modeled using \texttt{fleck}, with the spot contrast, spot radius, latitude, and longitude all set as free parameters with uniform priors, while the spot ellipsoid's sampling resolution is set independently. To capture baseline trends and instrumental systematics, quadratic polynomial coefficients ($c_0$, $c_1$, $c_2$), and a multiplicative scatter term ($scatter\_mult$) are fitted using Gaussian priors from white light curve fits, following default \eureka instructions. This configuration allows the white light curves to be modeled consistently in both transit geometry and stellar heterogeneity, producing the best-fit parameters summarized in \autoref{tab:fit_parameters}.

\begin{figure}[htbp]
  \centering

  \includegraphics[width=\columnwidth, trim={0.1cm 0cm 0cm 0cm}, clip]{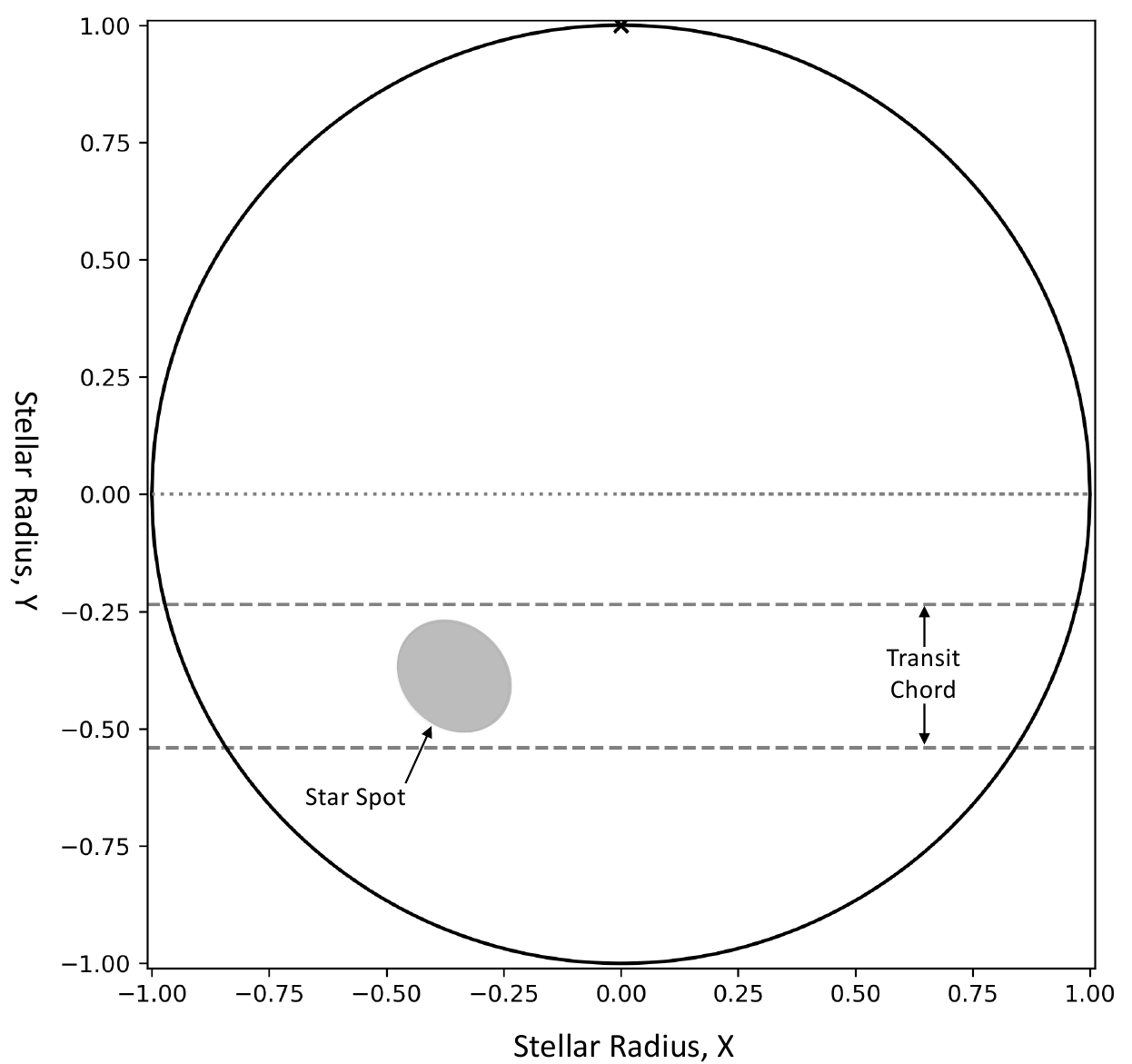}\\[\baselineskip]
  \includegraphics[width=\columnwidth, trim={0.1cm 0cm 0cm 0cm}, clip]{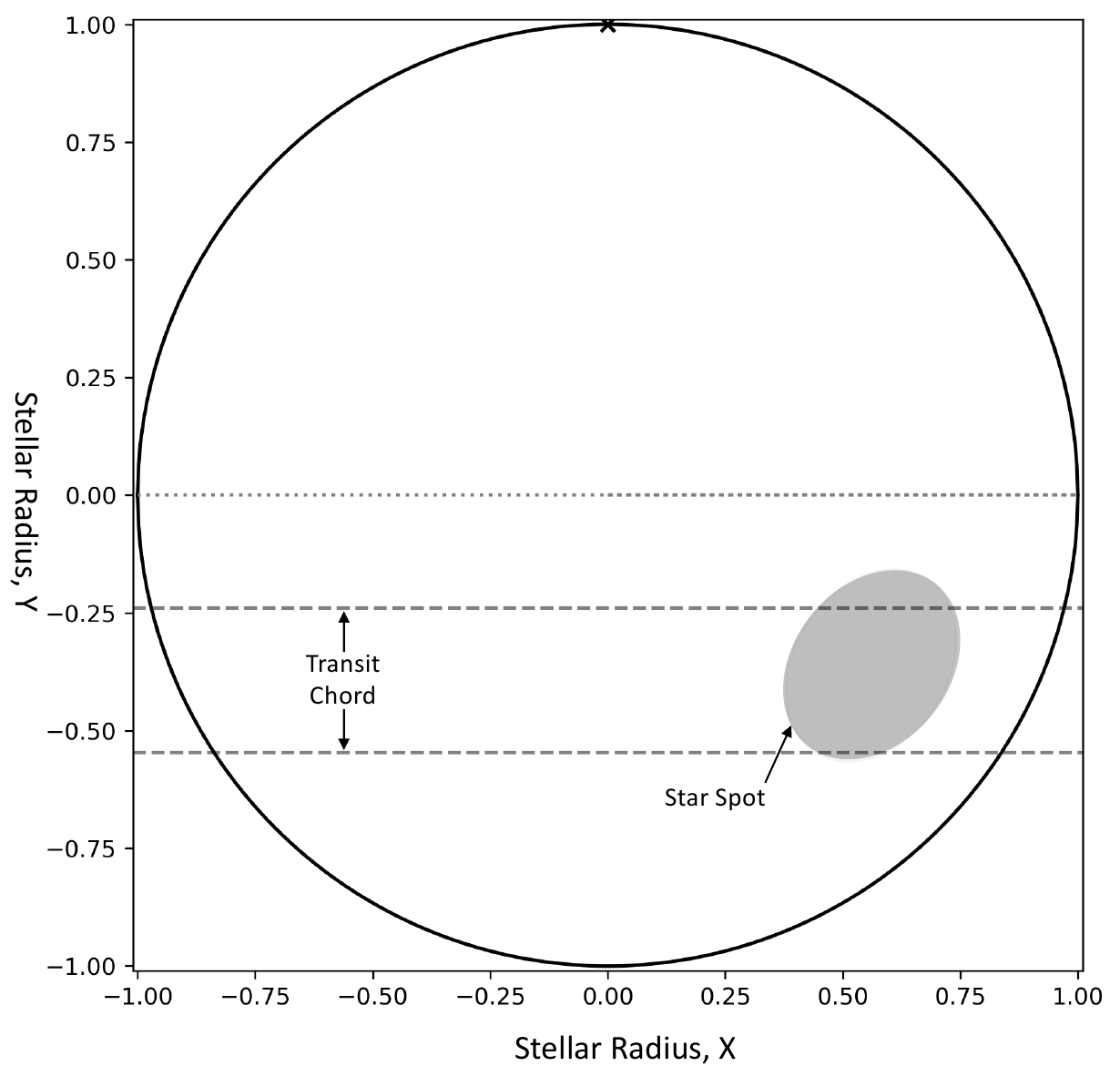}
  \caption{Modeled \fleck star spots for Visits 1 (top) and 2 (bottom). Visit 3 is not shown as the preferred transit model is spot-free for the \eureka\texttt{+ }\fleck reduction. Interestingly, the spots share roughly the same latitude across the transit chord, with similar spot contrasts of 0.960 $\pm$ 0.008 \& 0.971 $\pm$ 0.002 for Visits 1 and 2, respectively. Characteristics for the spots modeled for both reductions are detailed in \autoref{tab:fit_parameters}.}
  \label{fig:spots_fleck}
\end{figure}

The results of the white light curve fits for all three transits are shared in \autoref{fig:wlc_visits}. The first two transits are modeled with a single star spot, while the last transit prefers a spotless model. The residual distributions for all three visits are within 1$\sigma$. The \texttt{fleck} spots modeled for Visits 1 and 2 are shown in \autoref{fig:spots_fleck}. The final parameters from the best-fit white light curve models are listed in \autoref{tab:fit_parameters}.

Spectroscopic light curves for each visit were generated across 120 evenly binned (40 nm bins) channels, ranging from 0.6 to 5.3 $\mu m$. We fit each channel by adopting the best-fit system parameters derived from the white light curve analyses (\autoref{tab:fit_parameters}). The orbital period, eccentricity, and argument of periastron remain fixed to values from \cite{jordan_hats-74ab_2022}, the inclination and scaled semi-major axis are fixed to their best-fit white light curve values, while the planet-to-star radius ratio ($R_p/R_s$) is allowed to vary independently in each spectral channel. 

\begin{figure*}[t!]
    \centering
    \includegraphics[width=1.0\linewidth]{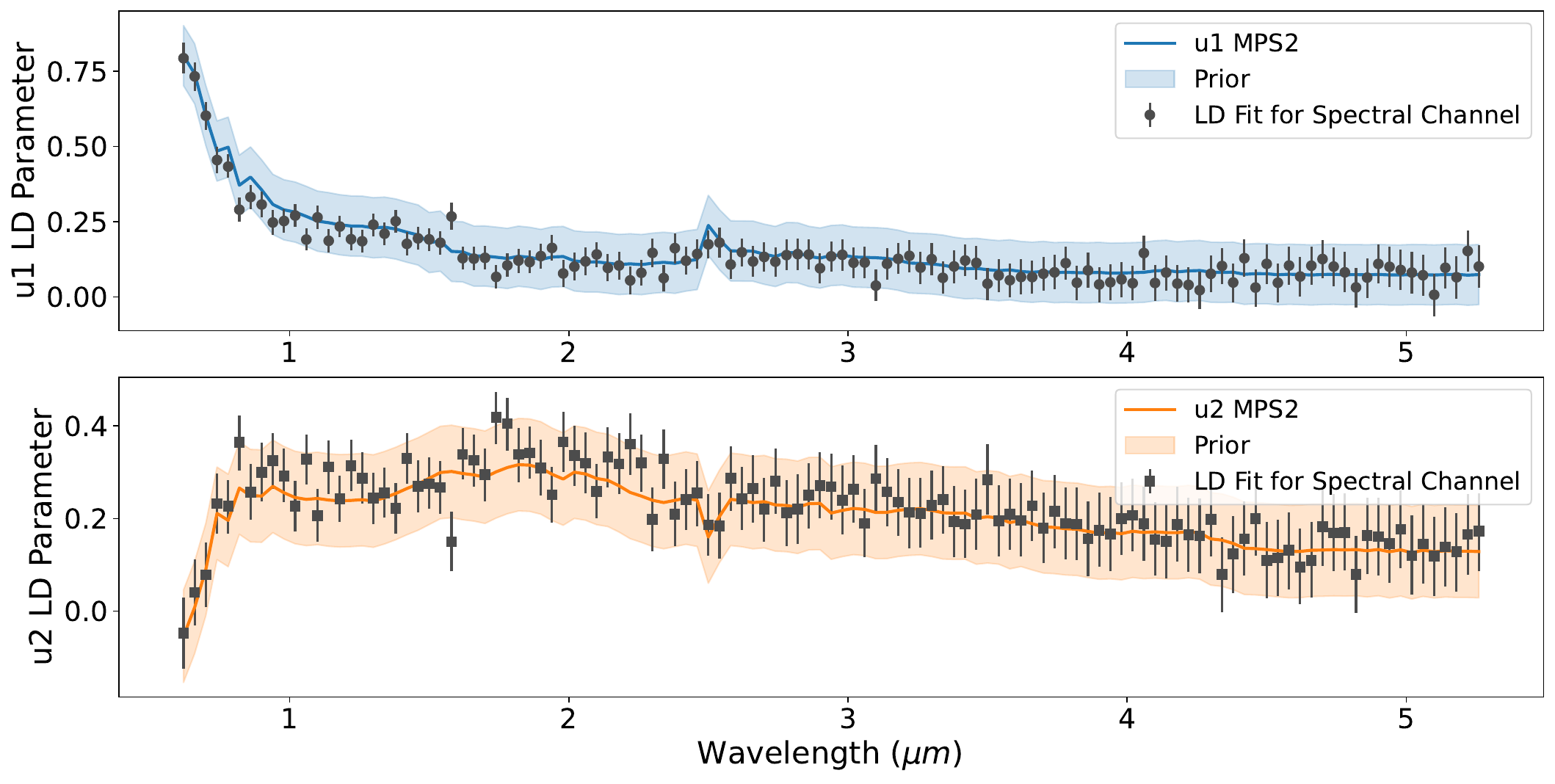} 
    \caption{Fitted spectroscopic limb darkening coefficients for a quadratic limb darkening law and posteriors against uniform priors centered about the \texttt{ExoTiC-LD} MPS2 stellar grid model (solid lines). Due to the agreement between the predicted and fit values, we fixed the limb darkening coefficients to the predicted values from the MPS2 grid when fitting the spectroscopic light curves.} 
    \label{fig:LD_analysis_mps2}
\end{figure*}

To determine the best stellar model for assessing limb darkening, we use the \texttt{recenter\_ld\_prior} feature in \eureka to fit limb darkening coefficients for a quadratic limb darkening law with Gaussian priors centered on the values predicted by a stellar model grid assuming a star with T$_\mathrm{eff}$ : 3790 K, log(g): 4.68, and Z: 0.5 dex \citep{Jordan2022}. Grid values are used as the initial parameter estimates and the mean in Gaussian prior with a width of 0.1. The best-fitting models using Phoenix, Stagger, MPS1, and MPS2 derived limb darkening coefficients were all evaluated, with reduced $\chi^2$ values of 1.24, 2.31, 1.15 and 1.02, respectively -- indicating that the limb darkening coefficients from the MPS2 stellar grid best-matched the data; where the $\chi^2$ values are averages for the spectral channel fits for both the $u1$ and $u2$ limb darkening coefficients. Based on this analysis shown in \autoref{fig:LD_analysis_mps2}, both data reductions use the \texttt{ExoTiC-LD} MPS2 stellar grid model for spectroscopic light curve fitting. 


The star spot contrast parameter is fitted freely across visits with a uniform prior between 0 and 1, with the spot geometry (radius, latitude, longitude) fixed to the best-fit solution from the respective visit white light curve. In both approaches, low-order polynomial coefficients (e.g., $c_0$, $c_1$, $c_2$) were fitted to account for systematics, and a multiplicative scatter term ($\kappa_{scatter}$) was included to mitigate residual white noise. This configuration ensured that only the spectrally dependent parameters (e.g., $R_p/R_s$, spot contrast, and baseline systematics) were fitted in the channel-by-channel analysis, while the orbital geometry and limb darkening were fixed using prior information.

\begin{figure*}[tb]
    \centering
    \includegraphics[width=1.0\linewidth, trim={0cm 0cm 0cm 0cm},clip]{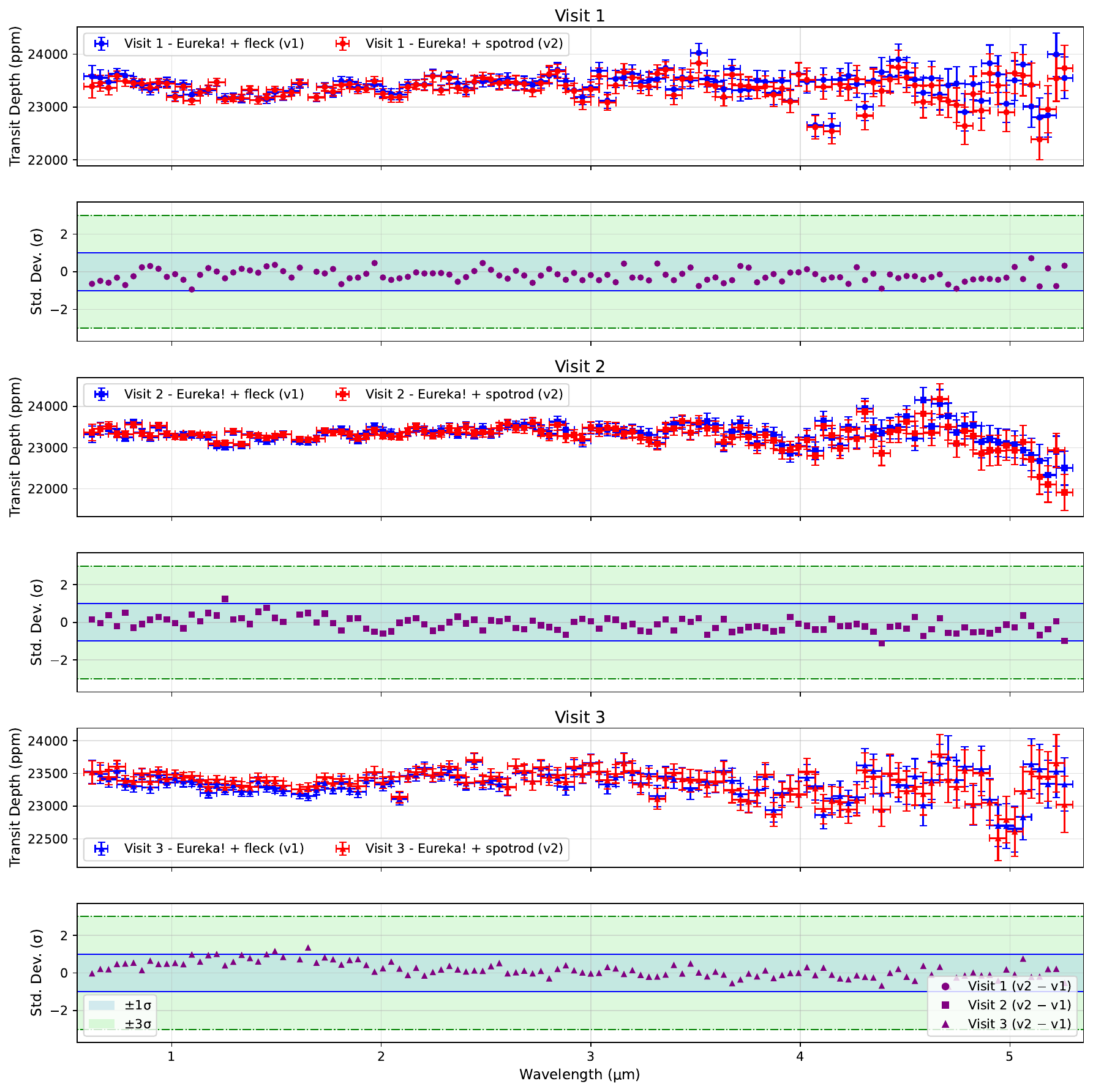} 
    \caption{
    A comparison of the independently-generated optimized \eureka\texttt{+ }\fleck and manual \eureka\texttt{+ }\spotrod transmission spectra for all three visits. Nearly all values fall within $\pm1\sigma$ relative to the other reduction, providing strong validation of the spectra generated for all visits across both approaches.}
    \label{fig:comparison_spectra}
\end{figure*}

\subsubsection{Manual Eureka! $+$ spotrod}
\label{subsec:data:reduction2} 

We verified the robustness of the optimized reductions by performing manual independent reductions of each visit with \texttt{Eureka!}. Starting with the \texttt{uncal} data, we reduced the data using \texttt{Eureka!} v1.1.2 and the \texttt{jwst} calibration pipeline v1.15.1 (CRDS pmap v1363) \cite{jwst.calibration}. Each visit shared the same \texttt{ecf} parameters which included an initial background subtraction and reference pixel subtraction utilizing the top and bottom 5 pixels in Stage 1, extracting the 1D stellar spectrum assuming a box-aperture width of 4-pixels centered on a Gaussian constructed profile of the 2D stellar spectrum, and then summing all wavelengths between 0.6 -- 5.3 $\mu$m to derive a white light curve. As each integration consisted of 5 groups, we opted to skip the jump step correction and masked detector pixel column 125, which demonstrated a significant spike in flux due to a hot pixel. The \texttt{ecf} files were not shared between the two independent reductions such that all the values chosen manually were completely uninformed of the optimized reduction performed in Section~\ref{subsec:data:reduction1}.

For each white light curve fit, we fixed the period, eccentricity and argument of periastron to the same values in \citet{Jordan2022}. The quadratic limb darkening coefficients were fitted assuming the \citet{kipping_efficient_ld_2013} re-parameterized values, which used uniform priors from 0 to 1. Similar to the \textit{Optimized Eureka! $+$ fleck} analysis, we fit for the $R_p/R_s$, inclination, scaled semi-major axis $a/R_s$, and mid-transit time. A second-degree polynomial was also used to further improve the RMS of the residuals.

We determined the number of star spots to include for each visit by first fitting a non-spotted transit model to the data using \texttt{spotrod} and \texttt{dynesty} with 5000 live points. We then created a time-average RMS curve of the residuals during transit and compared this with the residual points outside of transit. If the residual scatter in transit appeared significantly correlated compared to out-of-transit, we included a star spot to the model and re-fit. This process was repeated for each visit until in-transit and out-of-transit residuals showed similar near Gaussian noise. We found that two, one, and zero spots for each visit, respectively, minimized the in-transit residual scatter. The spot contrast, size, and location for each spot was fit independently between visits in which we did not account for possible stellar rotation. Spot contrasts were also allowed to float between individual spots with uniform priors spanning from a contrast of 0 (cold spots) to 1.2 (hot faculae).

A comparison of the best-fit white light curve parameters between the two reductions are listed in Table~\ref{tab:fit_parameters}. We find less than a 2$\sigma$ deviation in the planetary and orbital parameters between the two data reductions and subsequent analyses. However, there is significant deviation between the star spot parameters derived by \texttt{fleck} and \texttt{spotrod} with the latter preferring smaller spot sizes and darker spot contrasts. As discussed in \citet{murray.starspots}, spot parameters are degenerate, with multiple possible solutions yielding identical transit light curve models.

\begin{figure*}[]
    \centering
    \includegraphics[width=1.0\linewidth, trim={0cm 0cm 0cm 0cm},clip]{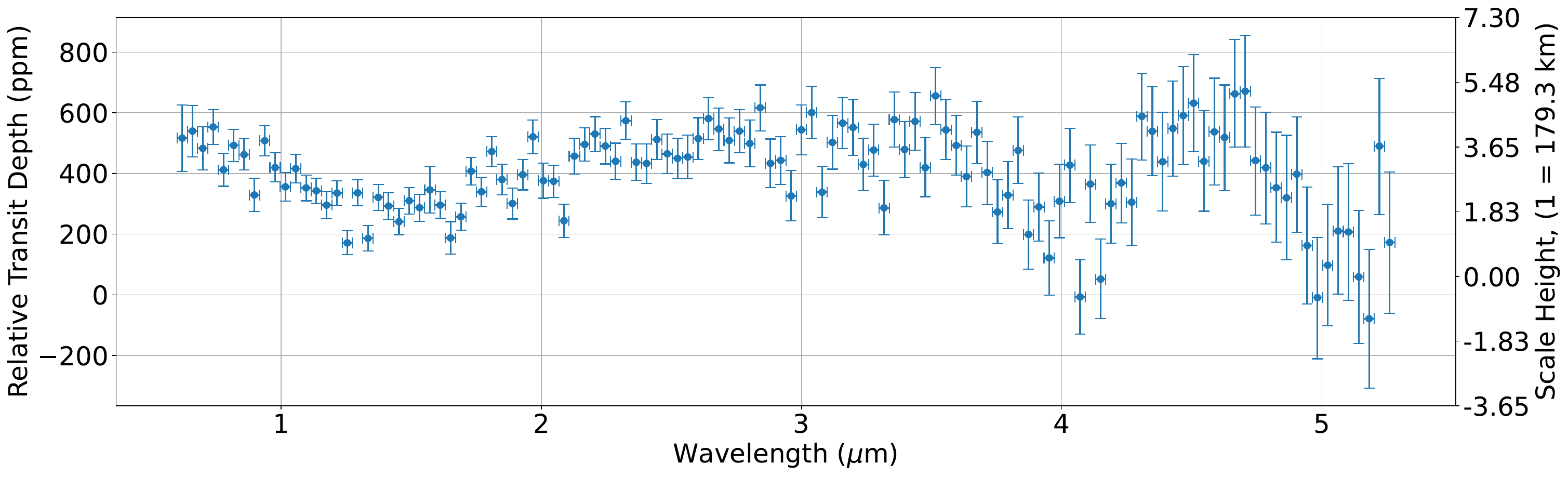} 
    \caption{
    The relative transit depth and scale heights for the optimized \eureka\texttt{+ }\fleck transmission spectra. The spectrum shown here is a weighted average of all three transits. Scale heights assume a mean molecular weight of $\mu =$ 2.2.}
    \label{fig:scale_height}
\end{figure*}

Spectroscopic light curves were generated using the same 120 wavelength bins, ranging from 0.6 to 5.3 $\mu m$. All parameters were fixed except for the radius ratio $R_p/R_s$, spot contrast, and the coefficients for a first-order linear polynomial. Limb darkening $u_1$ and $u_2$ were fit with a Gaussian prior centered on the \texttt{MPS2} derived values. Comparisons of the transmission spectra generated between the two reductions for each of the three transits are shown in \autoref{fig:comparison_spectra}. The two independently-derived datasets show a strong agreement of $\sim$ $1-\sigma$ between the spectra of all visits. With robust agreement between the two independent analyses of the spectra produced, a figure showing the weighted average spectrum (i.e. coadded spectrum) from all three visits derived by \textit{Optimized Eureka! $+$ fleck} is presented in \autoref{fig:scale_height}. 



\section{Atmospheric Retrievals} 
\label{sec:retrievals}

\subsection{Retrieval Setup}
\label{sec:retrievals:setup}

We use the \poseidon \citep[v1.2;][]{MacDonald2017, MacDonald2023, Mullens2024} atmospheric retrieval modeling suite to perform Bayesian inference on the \planetname{} transmission spectrum. We include opacities for the following species: \ce{H2O}, \ce{CH4}, \ce{CO2}, \ce{CO}, \ce{SO2}, \ce{H2S}, \ce{NH3}, \ce{HCN}, \ce{C2H2}, and \ce{PH3} \citep{Gordon2017HITRAN2016, HITRAN2020}. Collision-induced absorption (CIA) from HITRAN  \citep{Karman2019} is included for \ce{H2-H2}, \ce{H2-He}, \ce{H2-CH4}, \ce{CO2-H2}, \ce{CO2-CO2}, and \ce{CO2-CH4}. We resample opacities to $R=20,000$ in the NIRSpec PRISM wavelength range ($0.5 - 5.2$ \textmu{}m) and adopt an isothermal pressure-temperature profile. Clouds and hazes, when included, are parameterized using a gray cloud deck \citep{MacDonald2017} and a power-law haze slope \citep{LecavelierDesEtangs2008}. 

Our baseline retrievals use ``free chemistry'' to indicate that the $\log_{10}$ volume mixing ratio (VMR) of each molecule is included as a free parameter, assuming evenly mixed vertical gas abundances. \ce{H2} and \ce{He} are assumed to constitute the bulk composition of the atmosphere and are fractionated according to their relative cosmochemical abundances ($N_{\rm He}/N_{{\rm H}_2}=0.17$). We also use the ``chemical equilibrium'' mode in \poseidon to infer the atmospheric metallicity ([M/H]) and carbon-to-oxygen ratio (C/O) from the spectrum. \poseidon uses the code \texttt{FastChem} \citep{Stock2018} to compute equilibrium chemical abundances on a grid spanning the range $-1\le \log \mathrm{[M/H]} \le 4$ and $0.2 \le \mathrm{C/O} \le 2$. 

\poseidon also features a stellar contamination model to account for the transit light source (TLS) effect \citep{Rackham2017}, which we use to evaluate the necessity of a TLS component in our models. We fit for the effective temperature ($T$) and surface gravity ($\log g$) for the photosphere and any additional surface components included in the model. For such stellar heterogeneities, we fit for the fractional area covered by spots and faculae, $f_{\mathrm{spot}}$ and $f_{\mathrm{fac}}$ respectively, and assume that the remaining stellar area ($1-f_{\mathrm{spot}}-f_{\mathrm{fac}}$) is the photosphere. Our baseline \poseidon TLS contamination model includes two sources of heterogeneities from cool spots and hot faculae (and is henceforth referred to as the ``TLS2g'') and requires 8 additional free parameters (``8d'') in the model.    

Our baseline \poseidon retrievals also feature clouds and/or hazes, depending on the specific case. We adopted the gray cloud deck model from \citet{MacDonald2017} placed at the cloud-top pressure ($\rm \log P_{cloud}$) and we included patchy clouds ($\phi_{\mathrm{cloud}}$). For applicable models, we used a power law haze parameterized in terms of the Rayleigh enhancement factor relative to \ce{H2} Rayleigh scattering ($\log \rm a$) and the haze power law exponent ($\gamma$) \citep{MacDonald2017}.     

A full list of retrieval free parameters and their respective priors are listed in \autoref{tab:RetrievalResults}. 
Given the possibility of underestimated measurement uncertainties or model insufficiency that could present as systematics errors, we also included an error inflation term, $b$, that is parameterized following Equation 3 from \citet{Line2015} so as to add in quadrature with the measurement errors.  
All of our retrievals used the nested sampling code \texttt{MultiNest} \citep{Feroz2009} via \texttt{PyMultiNest} \citep{Buchner2014}. We used 2000 live points with a convergence criterion of $\Delta\ln Z=1$ which is sufficient to ensure robust sampling of the high-dimensional parameter space. We use several metrics to similarly quantify the retrieval results and compare model performance, including the Bayesian evidence ($\mathcal{Z}$), chi-squared ($\chi^2$) and reduced chi-squared ($\chi^2_{\nu}$) statistics. Bayes factors ($\mathcal{B}$) are used to compare nested models and, unless otherwise stated, indicate preference for the full model over a comparison model that omits a particular molecule or physical component (see \citealp{Trotta2008} for details and the adjectival descriptions adopted in this paper).   

\subsection{Retrieval Results}
\label{sec:retrievals:free}

\begin{figure*}[t]
    \centering
    \includegraphics[width=0.95\linewidth]{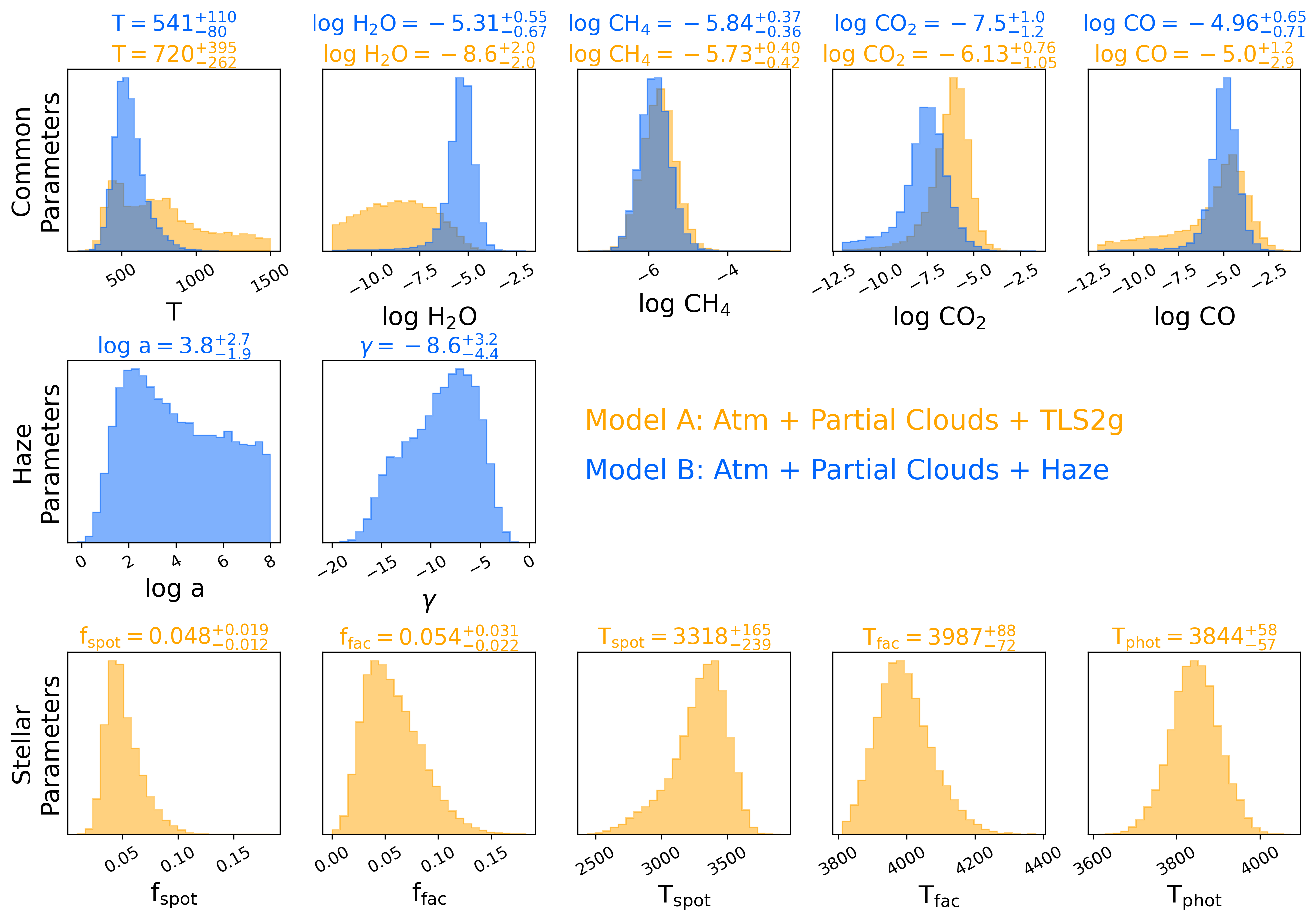} 
    \caption{Histograms of 1D marginalized posterior densities for a subset of parameters from the preferred TLS (orange) and hazy (blue) retrieval models. Despite two plausible retrieval solutions offered by Model A and Model B, both models provide similar constraints on the common atmospheric characteristics (top row). The primary difference resides in whether water is detected or not. }
    \label{fig:retrieval1_subplots}
\end{figure*}

\textit{Our free chemistry retrievals reveal two competing models that can similarly explain the \planetname{} transmission spectrum. The crux of the interpretation ambiguity lies in whether the slope at blue wavelengths ($<1.5$~\micron{}) is attributed to stellar contamination from unocculted heterogeneities on the stellar disk or a haze in the planet atmosphere, echoing similar findings to TOI-5205b \citep{Canas2025}. } \autoref{fig:retrieval1} highlights these two competing models in terms of their quality-of-fit to the three-visit coadded transmission spectrum and the impact of each physical and chemical model component on the observations. \autoref{fig:retrieval1_subplots} shows a subset of the 1D marginalized posteriors from these two retrieval models to compare parameters that are common to both models and to report constraints on parameters that are unique to each. \autoref{appendix:more_retrievals} provides full corner plots for both model solutions, including 1D and 2D marginalized posteriors for all free parameters in the fit. \autoref{tab:RetrievalResults} provides our numerical findings for both competing models, including $1\sigma$ posterior credibility intervals and $\chi^2_{\nu}$ and $\ln Z$.  We describe the two potential models here:  

\begin{figure*}[]
    \centering
    \includegraphics[width=0.85\linewidth]{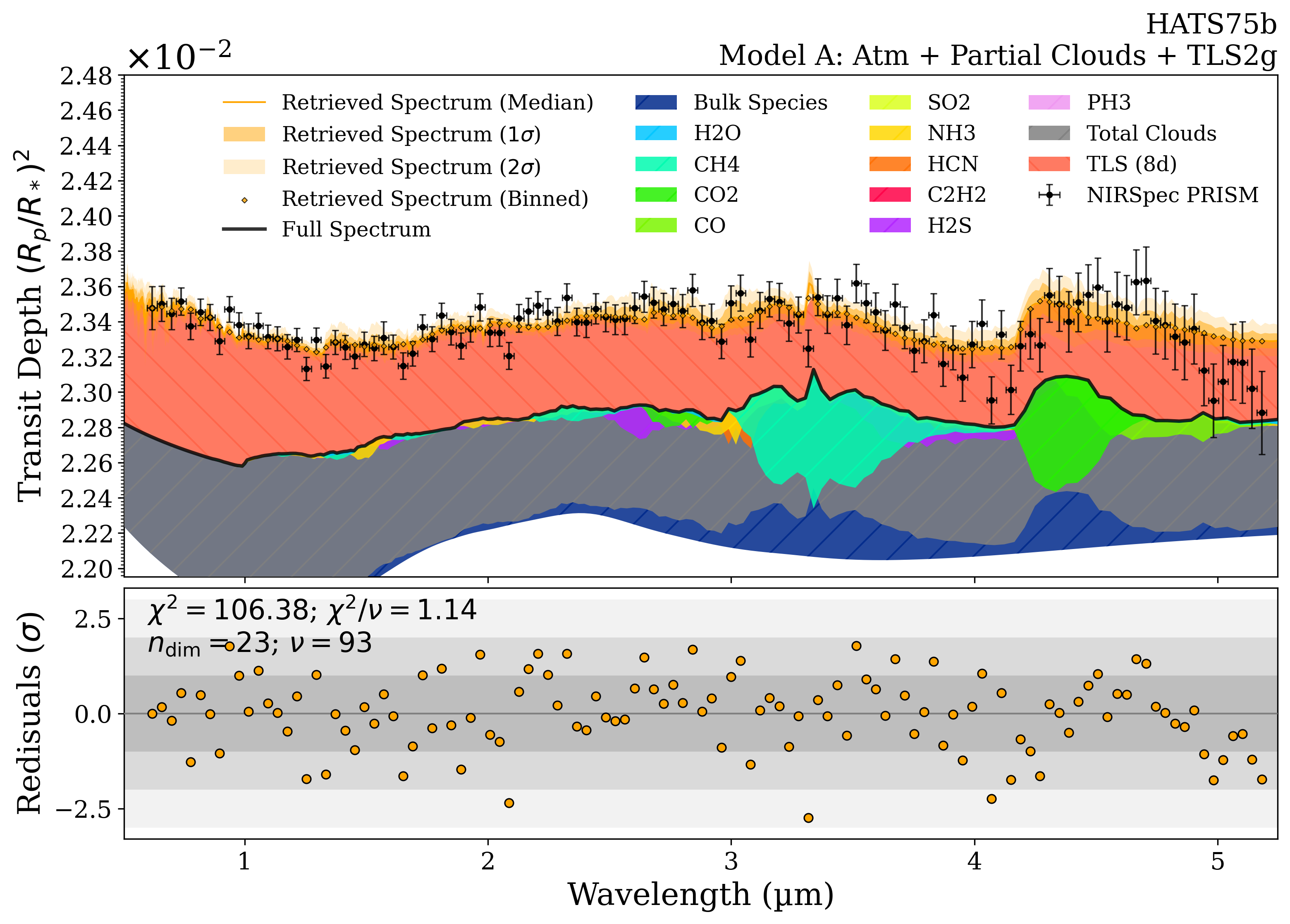} 
    \includegraphics[width=0.85\linewidth]{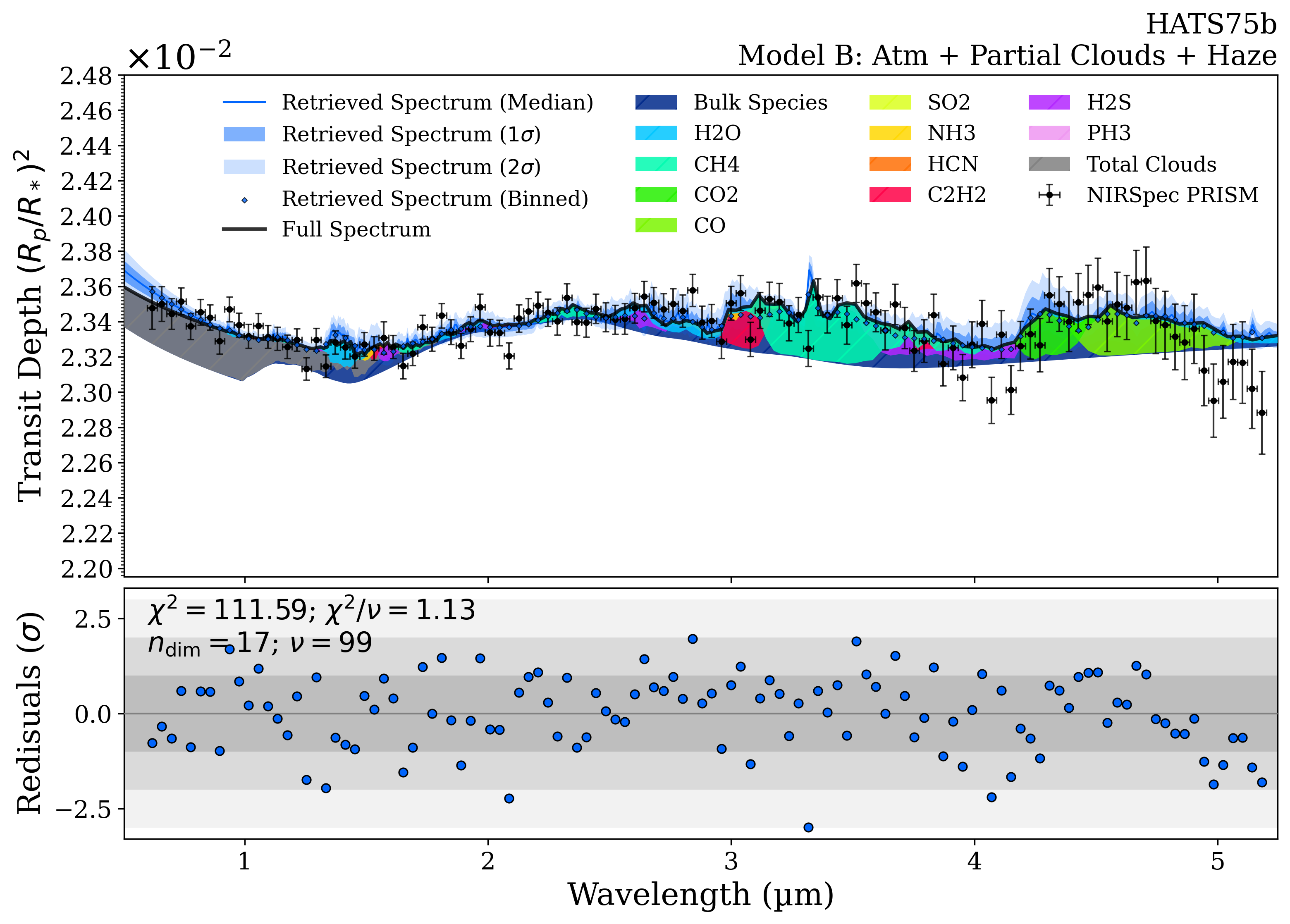} 
    \caption{Retrieval fits to the \planetname{} coadded transmission spectrum for the preferred TLS contamination model (top plot) and preferred haze model (bottom plot). Both plots show the spectral fit (upper panel) and the relative residual (lower panel). The data (black error bars) are compared against the median retrieved transmission spectrum with $1\sigma$ and $2\sigma$ credibility envelopes. The black model shows the best fitting planet spectrum simulated without TLS contamination (contribution highlighted in red). Shading below the best fit atmosphere shows molecular contributions to the spectrum. }
    \label{fig:retrieval1}
\end{figure*}

\begin{deluxetable*}{r|l||l|l}[t]
\tablewidth{0.97\textwidth}
\tablecaption{\poseidon Retrieval Model Free Parameters \& $1\sigma$ Posterior Constraints  \label{tab:RetrievalResults}} 
\tablehead{
\colhead{} & \colhead{} & \multicolumn{2}{c}{Posteriors} \\
\cline{3-4} 
\colhead{Parameters} & \colhead{Priors} & \colhead{\makecell[l]{Model A: Atm + Partial \\ Clouds + TLS2g}} & \colhead{\makecell[l]{Model B: Atm + Partial \\ Clouds + Haze}} 
}
\startdata
$\mathrm{R}_{\mathrm{p, \, ref}}$ [$\mathrm{R}_{\mathrm{J}}$]   & $\mathcal{U}(0.75, 1.02)$       & $0.847^{+0.012}_{-0.018}$      & $0.8614^{+0.0013}_{-0.0017}$ \\
$\mathrm{T}$ [K]                                                & $\mathcal{U}(200.00, 1500.00)$  & $720^{+394}_{-262}$            & $541^{+109}_{-80}$ \\
$\log \, \mathrm{H_2 O}$                                        & $\mathcal{U}(-12.00, -1.00)$    & $-8.6^{+2.0}_{-2.0}$           & $-5.31^{+0.55}_{-0.67}$ \\
$\log \, \mathrm{CH_4}$                                         & $\mathcal{U}(-12.00, -1.00)$    & $-5.73^{+0.40}_{-0.42}$        & $-5.84^{+0.37}_{-0.36}$ \\
$\log \, \mathrm{CO_2}$                                         & $\mathcal{U}(-12.00, -1.00)$    & $-6.13^{+0.76}_{-1.05}$        & $-7.5^{+1.0}_{-1.2}$ \\
$\log \, \mathrm{CO}$                                           & $\mathcal{U}(-12.00, -1.00)$    & $-5.0^{+1.1}_{-2.9}$           & $-4.96^{+0.65}_{-0.71}$ \\
$\log \, \mathrm{SO_2}$                                         & $\mathcal{U}(-12.00, -1.00)$    & $-9.5^{+1.6}_{-1.5}$           & $-9.7^{+1.6}_{-1.6}$ \\
$\log \, \mathrm{NH_3}$                                         & $\mathcal{U}(-12.00, -1.00)$    & $-8.6^{+1.8}_{-2.1}$           & $-7.7^{+1.4}_{-2.8}$ \\
$\log \, \mathrm{HCN}$                                          & $\mathcal{U}(-12.00, -1.00)$    & $-7.9^{+1.8}_{-2.3}$           & $-7.1^{+1.4}_{-3.0}$ \\
$\log \, \mathrm{C_2 H_2}$                                      & $\mathcal{U}(-12.00, -1.00)$    & $-8.0^{+1.3}_{-2.3}$           & $-7.8^{+1.1}_{-2.6}$ \\
$\log \, \mathrm{H_2 S}$                                        & $\mathcal{U}(-12.00, -1.00)$    & $-7.2^{+2.0}_{-2.9}$           & $-6.5^{+1.5}_{-3.5}$ \\
$\log \, \mathrm{PH_3}$                                         & $\mathcal{U}(-12.00, -1.00)$    & $-10.0^{+1.4}_{-1.2}$          & $-10.0^{+1.4}_{-1.3}$ \\
$\log \, \mathrm{P}_{\mathrm{cloud}}$ [bar]                     & $\mathcal{U}(-6.00, 2.00)$      & $-4.6^{+3.7}_{-0.9}$           & $0.52^{+0.99}_{-0.99}$ \\
$\phi_{\mathrm{cloud}}$                                         & $\mathcal{U}(0.00, 1.00)$       & $0.52^{+0.16}_{-0.22}$         & $0.54^{+0.29}_{-0.20}$ \\
$\log \, \mathrm{a}$                                            & $\mathcal{U}(-4.00, 8.00)$      & ---                            & $3.8^{+2.7}_{-1.9}$    \\
$\gamma$                                                        & $\mathcal{U}(-20.00, 2.00)$     & ---                            & $-8.6^{+3.2}_{-4.4}$   \\
$\mathrm{f}_{\mathrm{spot}}$                                    & $\mathcal{U}(0.00, 1.00)$       & $0.048^{+0.019}_{-0.012}$      & --- \\
$\mathrm{f}_{\mathrm{fac}}$                                     & $\mathcal{U}(0.00, 0.50)$       & $0.054^{+0.031}_{-0.022}$      & --- \\
$\mathrm{T}_{\mathrm{spot}}$ [K]                                & $\mathcal{U}(2300.00, 4574.40)$ & $3318^{+166}_{-241}$           & --- \\
$\mathrm{T}_{\mathrm{fac}}$ [K]                                 & $\mathcal{U}(3812.00, 5336.80)$ & $3987^{+88}_{-72}$             & --- \\
$\mathrm{T}_{\mathrm{phot}}$ [K]                                & $\mathcal{N}(3812.00, 79.00)$   & $3844^{+58}_{-57}$             & --- \\
$\log \, \mathrm{g}_{\mathrm{spot}}$ [cgs]                      & $\mathcal{U}(4.18, 5.18)$       & $4.62^{+0.30}_{-0.27}$         & --- \\
$\log \, \mathrm{g}_{\mathrm{fac}}$ [cgs]                       & $\mathcal{U}(4.18, 5.18)$       & $4.52^{+0.26}_{-0.21}$         & --- \\
$\log \, \mathrm{g}_{\mathrm{phot}}$ [cgs]                      & $\mathcal{U}(4.18, 5.18)$       & $4.97^{+0.13}_{-0.17}$         & --- \\
$\mathrm{b}$                                                    & $\mathcal{U}(-11.81, -5.28)$    & $-8.56^{+0.14}_{-0.14}$        & $-8.61^{+0.16}_{-0.19}$ \\
\hline
$\log~[\mathrm{M/H}]$ [dex] & --- & $-1.74^{+0.92}_{-0.76}$   & $-1.69^{+0.53}_{-0.45}$ \\
$\log~\mathrm{C/O}$ [dex] &   --- & $0.016^{+0.142}_{-0.039}$ & $-0.06^{+0.08}_{-0.21}$ \\
$\log~\mathrm{C/H}$ [dex] &   --- & $-1.49^{+0.96}_{-0.81}$   & $-1.53^{+0.60}_{-0.53}$ \\
$\log~\mathrm{O/H}$ [dex] &   --- & $-1.76^{+0.98}_{-0.92}$   & $-1.66^{+0.56}_{-0.52}$ \\
$\log~\mathrm{N/H}$ [dex] &   --- & $-3.1^{+1.1}_{-1.7}$      & $-2.4^{+0.7}_{-1.5}$ \\
$\log~\mathrm{S/H}$ [dex] &   --- & $-2.4^{+1.8}_{-2.0}$      & $-1.9^{+1.5}_{-2.4}$ \\
\hline
$\chi^2$                                    & ---                             & 106.38                         & 111.59   \\
$\ln \mathcal{Z}$                           & ---                             & 871.05                         & 875.99 \\
\enddata
\tablecomments{\textcolor{ForestGreen}{Gaseous species are specified in terms of their volume mixing ratio.} $\mathcal{U}$ and $\mathcal{N}$ denote uniform and normal prior distributions, respectively. We favor Model A due to the added knowledge on spots from resolved spot crossing events \autoref{sec:data:reduction}. The atmospheric metallicity ([M/H]) and elemental abundances (C/H, O/H, and S/H) are calculated from the free retrieval posteriors and are reported with respect to the solar values from \cite{lodders_solar_2020}. The adopted solar values are $[\mathrm{M/H}]_\odot=Z_\odot=0.011$, $\log~[\mathrm{C/H}]_\odot=-3.53$, $\log~[\mathrm{O/H}]_\odot=-3.27$, $\log~[\mathrm{N/H}]_\odot=-4.15$, and $\log~[\mathrm{S/H}]_\odot=-4.85$.} 
\end{deluxetable*}

\begin{itemize}
    \item TLS2g: 
A modest amount of stellar contamination from cool spots and hot faculae can explain the observed transmission spectrum (see \autoref{fig:retrieval1} upper panels) in combination with an atmosphere containing notable quantities of \ce{CO} ($\ln \mathcal{B} = 1.23$; ``weak evidence''), \ce{CH4} ($\ln \mathcal{B} = 8.25$; ``strong evidence''), and \ce{CO2} ($\ln \mathcal{B} = 3.13$; ``moderate evidence'') (see \autoref{fig:retrieval1_subplots} upper panels). The preferred TLS solution contains approximately 5\% spots and 5\% faculae covering fractions. The isothermal temperature is poorly constrained at $\rm T=720^{+394}_{-262}~K$ due to broad cloud parameters that tend to favor high altitude (low pressure) clouds and allow for deeper clouds at lower planetary temperatures. We acknowledge that using isothermal temperatures can cause retrieval biases \citep[][]{Rocchetto2016, Barstow2020}, but our initial testing did not suggest any sensitivity of our results to vertical structure in the PT profile. Rather, degeneracies between several parameters due to continuum normalization \citep[e.g., see][]{heng2017theory, fisher2018retrieval} are exacerbated when TLS and clouds are both present, which renders the retrieved atmospheric temperature highly uncertain. While \ce{CO}, \ce{CH4}, and \ce{CO2} are well constrained, we do not detect \ce{H2O} ($\ln \mathcal{B} = 0.18$; ``inconclusive'') and instead obtain an upper limit on \ce{H2O} at $\rm \log(VMR)=-5.41$ dex ($2 \sigma$). 
    \item Haze: 
Alternatively, an atmosphere with a haze that produces a scattering slope $\le1.2$ \microns{} can instead explain the observed \planetname{} transmission spectrum (see \autoref{fig:retrieval1} lower panels). In this hazy case, we again detect the spectrally active gases \ce{CO} (``weak evidence''), \ce{CH4} (``strong evidence''), and \ce{CO2} (``moderate evidence''), and now also detect \ce{H2O} ($\ln \mathcal{B} = 1.93$; ``weak evidence'') with $\rm \log(VMR)= -5.31^{+0.55}_{-0.67}$ dex ($1 \sigma$). The haze parameters, $\rm \log a$ and $\gamma$, are poorly constrained because they are correlated with each other and the fractional cloud coverage. 
\end{itemize}

Despite the two different solutions for the blue slope in the transmission spectrum, both the TLS and hazy cases produce otherwise similar findings for the atmospheric composition of \planetname. The retrieved isothermal temperature and trace gas abundances are similar and within $1\sigma$. The primary difference, in addition to the inclusion of TLS or haze, is that the hazy model places a lower limit on \ce{H2O}, thereby weakly detecting it, whereas the TLS model does not require \ce{H2O}, likely due to the presence of \ce{H2O} in the unocculted spots \citep{Moran2023}. In addition, the hazy model favors deep clouds, while the TLS model favors high-altitude clouds. 

We ran additional models with more and less complex retrieval setups to assess the levers that drive these two solutions. We found that a maximally complex retrieval that simultaneously fits the haze and the TLS (via the 8d \texttt{two\_spots\_free\_log\_g} model) acts to minimize the haze and reproduces the aforementioned TLS solution, including the \ce{H2O} non-detection (with $1\sigma$ upper limit of about 1 ppm). However, when we reduce the complexity of the TLS model to the \texttt{one\_spot} model (with only 3 dimensions), the TLS solution is suppressed and the aforementioned hazy solution emerges. We conclude that the hot faculae component of the TLS contamination model plays an important role in our TLS explanation, but if a hazy atmosphere is invoked, TLS becomes an unnecessary addition to the model. 

It is non-trivial to conclude that one of the two models should be favored over the other. The hazy model yields a marginally better goodness-of-fit than the TLS model with $\chi^2/\nu=1.13$ and $\chi^2/\nu=1.14$, respectively. The hazy model also has a higher Bayesian evidence than the TLS model with $\rm \ln \mathcal{Z} = 875.99$ and $\rm \ln \mathcal{Z} = 871.05$, respectively. Based on these numbers alone, the hazy model should be the preferred model. However, this does not imply that the hazy model is more physically correct, but rather that the additional parameters of the TLS model (6 additional free parameters) cannot be justified since the data can be nearly as well fitted by the simpler model. In light of additional contextual physical clues from the spot crossings, the TLS model should not be so readily discounted. That is, we observed clearly evident star spot crossing events in the white light curve (\autoref{fig:wlc_visits} and \autoref{fig:spots_fleck}) and their reported contrasts (see \autoref{tab:fit_parameters}) are similar to the spots found contaminating the transmission spectrum. We also used the internal \poseidon stellar modeling framework to fit the out-of-transit calibrated stellar spectrum. All three visits yield highly consistent results; Visit 3 has the average (middle) result with: $\rm T_{phot} = 3767 \pm 74~K$, $\rm T_{spot} = 3578^{+107}_{-287}~K$, $\rm T_{fac} = 3909^{+156}_{-82}~K$, $\rm f_{spot} = 0.23^{+0.30}_{-0.20}$, and $\rm f_{fac} = 0.25^{+0.31}_{-0.20}$. Although the disk-integrated stellar fits poorly constrain the spot and facula fractions compared to the stark sensitivity of the transmission spectrum to stellar heterogeneities, these two independent constraints on the stellar properties are consistent within $1\sigma$. Both approaches allow for the ${\sim}5\%$ covering fractions of both spots and faculae. Nonetheless, while our results from the disk-integrated spectrum are consistent with our TLS solution, they are ultimately inconclusive when it comes to the hazy solution because the spot and faculae coverage are so poorly constrained that we gain no insight into favoring or disfavoring the hazy solution. We caution that just because the two-parameter power law haze model may yield the simplest viable retrieval solution, does not mean that it presents a more representative physical interpretation than the complex 8-parameter TLS model. Therefore, while haze may be the simplest explanation in terms of model complexity and favored in terms of statistical evidence, TLS contamination in the transmission spectrum is supported through other independent lines of evidence and is our adopted model.

Our free retrievals suggest a low sub-solar metallicity atmosphere with a high super-solar C/O ratio. We post-processed the volume mixing ratio posteriors from our free retrievals to determine posteriors for the gas phase atmospheric metallicity and C/O ratio. Our retrieval with TLS result in $\rm log~[M/H]=-1.74^{+0.92}_{-0.76}$ and $\rm C/O=1.04^{+0.40}_{-0.09}$. Our retrieval with haze result in $\rm log~[M/H]=-1.69^{+0.53}_{-0.45}$ and $\rm C/O=0.87^{+0.17}_{-0.33}$. As with the retrieved VMRs shown in \autoref{fig:retrieval1_subplots} and \autoref{tab:RetrievalResults}, the metallicity and C/O are quite consistent regardless of the ambiguity between TLS contamination and haze. Carbon monoxide is the strongest driver of constraints on the C/O and [M/H] due to its relatively high abundance. Whereas the difference in C/O between the TLS and hazy models is primarily due to the lower limit on \ce{H2O} in the hazy case, which drives oxygen up and C/O down.   

In addition, we ran variants on the two preferred models using \poseidon in equilibrium chemistry mode, but encountered model limitations due to the low metallicity\footnote{The chemical equilibrium retrievals initially favored an erroneous high metallicity solution that provided a poor fit to the data. When forced to consider only sub-solar metallcities (with the restrictive prior $\rm \log Met \sim [-1,0]$), the retrievals identified a better fitting solution that pressed against the lower bound of the metallicity prior set by limitations of the precomputed \texttt{FastChem} grid.}. Therefore, while there could be additional molecules in the atmosphere that remain unseen and that would increase the metallicity, we conclude that  \planetname{} very likely has a sub-solar atmospheric metallicity.  

\section{Bulk Metallicity}\label{sec:bulk}

Similar to the previous result from this survey ---TOI-5205b \citep{Canas2025} --- we used the measured atmospheric metallicity, planetary bulk properties and stellar age to put preliminary constraints on the total planetary metallicity (Z$_{planet}$) of \planetname{}. To determine the bulk metallicity of HATS-75b, we used the open-source thermal evolution model \texttt{GASTLI} \citep{acuna_gastli_2024} and a Markov Chain Monte Carlo approach. \texttt{GASTLI} builds a set of hydrostatic models at specific intrinsic temperatures and generates a cooling track by integrating the luminosity equation along the hydrostatic models.

We first created a grid of cooling tracks varying the planetary mass, atmospheric metallicity, core mass fraction (CMF), and intrinsic temperature. The tabulated atmospheres in \texttt{GASTLI} were limited to a lower value of log(M/H) = -2, which therefore is the lower bound of our models. Since the high inferred C/O fraction is beyond the upper limit of the atmospheric models in \texttt{GASTLI}, we set it to the maximum value of 0.55.

We used the system age as an independent variable to match the observed planetary mass and radius. This required an additional pre-processing step to map the cooling tacks calculated by \texttt{GASTLI} onto a regular age grid, and was achieved by creating and using a one-dimensional interpolator built from the calculated ages and the grid of intrinsic temperatures.

A linear four-dimensional interpolator on the grid of evolution models served as the forward model. We first used an input grid in 4D using uninformed uniform priors for the mass (from 0.24 to 0.70 M$_{\rm{J}}$), core mass fraction (0 to 0.7), atmospheric metallicity log[M/H] (from -2 to 1), and age (1 to 10 Gyr), with planet radius and intrinsic temperatures as the outputs. For the bulk-metallicity retrieval, we then employed a Markov-Chain Monte Carlo approach and used the forward model to calculate the planet radius and intrinsic temperature. The statistical model was constrained by three Gaussian distributions representing the observed planetary mass, radius, and the retrieved atmospheric metallicity.

\begin{figure*}[t]
    \centering
\includegraphics[width=0.9\linewidth]{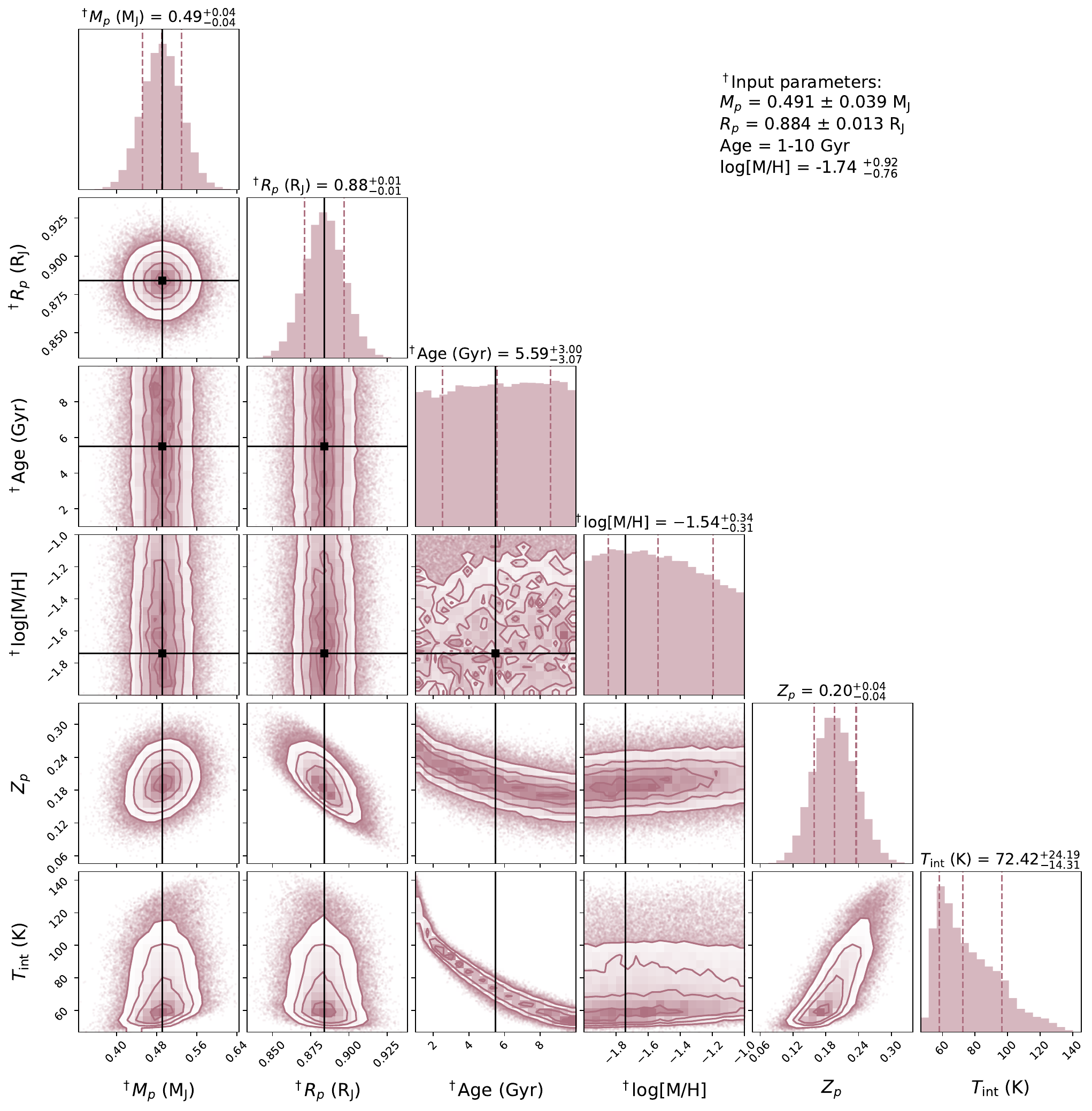} 
    \caption{Estimates for bulk metallicity $Z_{\rm{p}}$ and the intrinsic temperature $T_{\rm{int}}$ using measurements of planet mass, radius, atmospheric metallicity, and the system age. The input parameters for the statistical model are marked with a $^\dagger$ and listed in the figure.}
    \label{fig:z_bulk_corner}
\end{figure*}

The results are shown in \autoref{fig:z_bulk_corner}. We first note our models were able to match all observational constraints (atmospheric metallicity, planetary mass and radius) well within their uncertainties. Unsurprisingly, we are not able to constrain the age using this framework, given the slow cooling beyond 1 Gyr. However, this is consistent with measured rotation period from \cite{jordan_hats-74ab_2022} of $\sim$35 days, which suggests an age around 4 Gyr based on the age-rotation relations from \cite{engle_living_2023-1}, albeit with significant uncertainty. Importantly, the inferred bulk metallicity of HATS-75b ($Z_p \simeq$ CMF, since $Z_{\rm{atm}}$ is small) is found to be $Z_p$ = 0.20 $\pm$ 0.04. This is higher than earlier results from \citet{2025A&A...693L...4M} that are based on the \texttt{planetsynth} \citep{2021MNRAS.507.2094M} evolution models and used an older stellar age estimate. Crucially, the value differs sharply from the very low measured atmospheric metallicity of log[M/H] $\sim$ -2 (i.e., $\sim$1$\%$ solar; $Z_{\rm atm} \sim 10^{-4}$ as a metal mass fraction). \textit{This strongly implies that the planet cannot be fully mixed.}

Although the approach used in this work benefits from atmospheric constraints that reduce some of the degeneracies in understanding the planet's composition \citep{muller_warm_2023, muller_towards_2023}, there are still caveats worth noting. As we have learned from detailed \textit{in-situ} observations of the Solar System giant planets, giant planet interiors can be remarkably complex, with phase transitions causing regions of H-He immiscibility \citep{mankovich_evidence_2020, sur_simultaneous_2024, howard_evolution_2024, bodenheimer_formation_2025}, diluted fuzzy cores with metallicity gradients \citep{helled_fuzziness_2017,vazan_jupiters_2018,muller_challenge_2020}, inverted heavy-element gradients in the outer envelope \citep{debras_new_2019,howard_exploring_2023, muller_can_2024} and more \citep{helled_nature_2022, miguel_jupiters_2022}. Furthermore, there are limitations to the equations of state and energy transport assumptions \citep[see][for further discussion]{Canas2025}. We therefore emphasize that Fig. \ref{fig:z_bulk_corner} shows the formal uncertainties from the MCMC retrieval; the true theoretical uncertainties are likely much larger. However, the purpose of this JWST survey is to attempt and marginalize over these caveats by performing a homogeneous analysis of the bulk metallicity for the planets in this sample.

\section{Discussion}
\label{sec:discuss}

The NIRSpec/PRISM transmission spectrum of \planetname{} presents an intriguing ambiguity that echoes patterns seen in other GEMS \citep{Canas2025}. As illustrated in \autoref{fig:retrieval1}, our retrievals yield two comparably good explanations for the observed spectral features: one invokes unocculted stellar heterogeneities (TLS effect) while the other attributes the blue-ward spectral slope to a high-altitude haze in the planet’s atmosphere. Both models successfully reproduce the prominent absorption signatures of CO, CH$_4$, and CO$_2$ in the NIRSpec/PRISM bandpass, including consistent abundances between the two models. The key difference lies in the origin of the rising slope at $\lambda \lesssim 1.5~\mu$m and the presence (or absence) of H$_2$O. In the stellar contamination scenario, the slope is explained by a modest coverage of starspots and faculae on the M-dwarf host, whereas in the hazy atmosphere scenario, the slope is intrinsic to the planet, arising from Rayleigh-like scattering by airborne haze particles that also permit a detection of water vapor. \autoref{fig:retrieval1_subplots} highlights how these two interpretations diverge in fitted parameters. Notably, the haze model retrieves a significant H$_2$O abundance, while the TLS model is consistent with an upper limit on H$_2$O, with the water absorption potentially being masked by spot-induced spectral contamination.

We note that the TLS and haze interpretations are not mutually exclusive, and that both effects could contribute to the observed short-wavelength slope. Although haze formation is expected in this temperature regime, several factors favor a dominant contribution from stellar contamination: (1) a photometric measurement of the stellar rotation period induced by spot modulation \citep{jordan_hats-74ab_2022}, (2) resolved spot-crossing events in the white light curves (\autoref{fig:spots_fleck}), and (3) the low-metallicity atmosphere \citep{canas_gems_2025}. Low-metallicity atmospheres are disfavored for haze formation relative to high-metallicity atmospheres, as reduced heavy element abundances limit the availability of condensates or photochemical precursor species \citep{canas_gems_2025, crossfield_mapping_2025}. This being said, the H$_2$O abundance in the TLS retrieval is comparable to that of the haze scenario, as the corresponding posteriors exhibit similar constraints (\autoref{fig:retrieval1_subplots}).

\textit{Considering the broader physical context, we favor the stellar heterogeneity interpretation as the more plausible explanation for \planetname{}'s spectrum.} Empirical precedent for TLS-dominated spectra is growing among M-dwarf planets.  For example, TOI-5205b from this survey showed a similarly steep optical slope that was ultimately attributed to unocculted starspots rather than to the planet’s atmosphere. In our case, the white-light curve analysis already revealed in-transit spot-crossings (see \autoref{fig:wlc_visits} and \autoref{fig:spots_fleck}), indicating that HATS-75’s photosphere is indeed spotted. Our retrieval finds that a combination of $\sim$5\% cool spots (at $T_{\rm spot}\approx 3320$ K) and a comparable fractional coverage of warm faculae ($T_{\rm fac}\approx 4000$ K) on the stellar disk with a photosphere temperature of $\rm 3844 \pm 58~K$ can fully explain the observed blueward slope in \planetname{}'s spectrum. These stellar properties are consistent with the star’s previously reported effective temperature ($\rm T_{\rm eff} = 3790 \pm 6$ K; \citealp{Jordan2022}), suggesting an inhomogeneous photosphere with temperature contrasts of a few hundred Kelvin. Under this interpretation, the planetary spectrum is mostly free of obscuring haze and the muted water features are explained by the fact that cool starspots imprint water absorption features in the transmission spectrum, diluting or mimicking the planetary H$_2$O signal. Such a mechanism is well documented for M-dwarfs: excess water vapor in unocculted starspots relative to the photosphere can mimic planetary atmospheric water features (e.g. H$_2$O vs. stellar contamination in \citealt{Moran2023}). As a result, we caution that the sub-solar water abundance retrieved in the TLS scenario may not reflect the true atmospheric composition of \planetname, but rather the influence of stellar contamination. The transmission spectra of GEMS require careful treatment of stellar heterogeneity and activity. 

Importantly, the spot crossing events provide an independent, wavelength-resolved constraint on the stellar heterogeneity. \autoref{fig:spotcontrasts} demonstrates this, with the measured spot contrasts from the two spotted visits closely following the spectral structure predicted by retrieved stellar atmosphere models, despite an overall offset in the absolute contrast level relative to the retrieved spot temperature. This agreement in spectral shape suggests both the spot-crossing analysis and the atmospheric retrievals are probing the same underlying photospheric inhomogeneities, with residual differences in inferred temperatures arising from degeneracies between spot coverage, spot temperature, and stellar limb-modeling.

An alternative explanation is that \planetname{}'s atmosphere possesses a high-altitude aerosol layer responsible for the optical slope. If the spectral slope is due to planetary haze, our retrieval indicates the haze must be quite opaque at pressures lower than ${\sim}0.3$ bar and potentially composed of small particles ($\gamma \approx -13$ to $-5.6$ at $1\sigma$) to produce the observed scattering signature. The hazy atmosphere scenario yields a statistically comparable fit to the data (slightly higher Bayesian evidence and a marginally lower reduced $\chi^2$ than the TLS model) and, importantly, it finds H$_2$O as a detectable constituent of the atmosphere ($X_{\rm H_2O}\sim10^{-5}$), alongside CH$_4$, CO, and CO$_2$. The presence of CH$_4$ at this temperature, together with the high retrieved C/O ratio (C/O $\approx0.9$–1.0), suggests that \planetname{}'s atmosphere could be relatively carbon-rich and oxygen-poor. 

\textit{While the haze interpretation provides a satisfying explanation for the clear detection of H$_2$O in the spectrum, we regard it as less physically-favored given the indications of stellar activity (see \autoref{fig:spots_fleck})}. The primary concern being that the haze model possibly compensates for unmodeled stellar effects, yielding a degeneracy that highlights the importance of stellar characterization alongside transmission spectrum analyses. 


Breaking this degeneracy between stellar heterogeneity and atmospheric haze will be crucial for confidently interpreting \planetname{}'s atmosphere as well as other similar transiting planets. Future observations with broader wavelength coverage offer a promising path forward. In particular, JWST/MIRI LRS could extend the transmission spectrum into the mid-infrared (5–12 $\mu$m), where the influence of unocculted starspots is expected to diminish and molecular \citep{Moran2023} or aerosol absorption features \citep{Grant2023} become more pronounced. A JWST/MIRI emission spectrum would mitigate the impact of stellar contamination and offer an independent measurement of the \planetname{} atmosphere free from the TLS effect. 

Ultimately, distinguishing between these two scenarios for \planetname{} has far-reaching implications, affecting the inferred atmospheric composition, and thus influencing theoretical interpretation of this planet’s formation and evolution. As one of the rare giant planets around an M-dwarf, \planetname{} underscores both the opportunities and challenges of atmospheric characterization in this regime, offering a valuable case study where star and planet must be studied together to reveal the planet’s true nature.

\section{Conclusion} 
\label{sec:conclusion}

The JWST/NIRSpec transmission spectrum of \planetname{} is most consistently explained by stellar contamination due to TLS, supported by observed star spot crossings and modest faculae coverage required to reproduce the spectral slope. Our retrievals under this model indicate strong evidence for CH$_4$ ($\ln \mathcal{B} = 8.25$), moderate evidence for CO$_2$ ($\ln \mathcal{B} = 3.13$), weak evidence for CO ($\ln \mathcal{B} = 1.23$), and a tentative non-detection of H$_2$O due to spot-induced degeneracy. The atmosphere is consistent with sub-solar metallicity and elevated C/O. While a haze-dominated model can fit the data equally well, we disfavor the hazy scenario based on independent circumstantial evidence provided by star spot crossings that reveal characteristics similar to those required to produce the observed TLS contamination. Regardless of the remaining model ambiguity, both prescriptions result in an atmosphere that is metal-poor $<$ -1.5 dex) and carbon rich. When coupled with simplified interior models of the planet's bulk, these predict a bulk metallicity that is $\sim$ 2 dex more metal rich than the atmosphere, again indicating the lack of convective mixing in this warm GEMS.

\planetname{} thus joins the rare population of giant planets around M-dwarfs observed with JWST, underscoring both the challenges of stellar contamination in transmission spectroscopy and the importance of careful host-star treatment in revealing the true chemistry of these uncommon worlds.

\newpage

\begin{acknowledgments}

This work is based on observations made with the NASA/ESA/CSA James Webb Space Telescope. The data were obtained from MAST at STScI, which is operated by the Association of Universities for Research in Astronomy, Inc., under NASA contract NAS 5-03127 for JWST. These observations are associated with program \#3171. Support for program \#3171 was provided by NASA through a grant from the Space Telescope Science Institute, which is operated by the Association of Universities for Research in Astronomy, Inc., under NASA contract NAS 5-03127.

The JWST data presented in this paper were obtained from MAST at STScI. The specific observations analyzed can be accessed via \dataset[DOI: 10.17909/dtza-4154]{https://doi.org/10.17909/dtza-4154}. Support for MAST for non-HST data is provided by the NASA Office of Space Science via grant NNX09AF08G and by other grants and contracts.

This research has used the NASA/IPAC Infrared Science Archive, which is funded by the National Aeronautics and Space Administration and operated by the California Institute of Technology.

AI-assisted tools (ChatGPT, OpenAI) were used for editorial support, refining LaTeX structure, and providing debugging guidance for \texttt{python} code. These tools did not contribute to any scientific analysis, data interpretation, or the generation of scientific results. All scientific content, methodology, and conclusions are the responsibility of the authors.

CIC acknowledges support by (i) NASA Headquarters through an appointment to the NASA Postdoctoral Program at the Goddard Space Flight Center, administered by ORAU through a contract with NASA and (ii) NASA under award number 80GSFC24M0006.

The Center for Exoplanets and Habitable Worlds is supported by the Pennsylvania State University and the Eberly College of Science.

\end{acknowledgments}

\software{
\texttt{Eureka!} \citep{Bell2022};
\texttt{ExoTiC-LD} \citep{exotic.ld}; 
\texttt{fleck} \citep{Morris2022};
\texttt{spotrod} \citep{Beky2014};
\texttt{DYNESTY} \citep{Speagle2020};
\texttt{GASTLI} \citep{acuna_gastli_2024};
\texttt{PySynPhot} \citep{pysynphot};
\texttt{MultiNest} \citep{Feroz2009};
\poseidon \citep{MacDonald2017, MacDonald2023};
\texttt{PyMultiNest} \citep{Buchner2014};
\texttt{astropy}\citep{2013A&A...558A..33A,astropy_collaboration_astropy_2018,astropy_collaboration_astropy_2022}}

\facilities{JWST(NIRSpec), MAST, ADS}

\clearpage
\appendix
\setcounter{figure}{0}
\renewcommand{\thefigure}{A\arabic{figure}}
\setcounter{table}{0}
\renewcommand{\thetable}{A\arabic{table}}

\onecolumngrid

\section{Additional Data Analysis Materials} 
\label{appendix:more_data_analysis}

\noindent This appendix contains supporting material for the data reduction described in \autoref{subsec:data:reduction1}. 

For the \textit{optimized} \eureka\ dataset, we include the output of the \texttt{Eureka!} optimizer used to guide parameter selection for Stages~1, 3, and 4 of \texttt{Eureka!}. The optimizer sequentially adjusts configuration values such as background subtraction, ramp fitting, and outlier rejection thresholds, evaluating each trial using the MAD value of the resulting spectroscopic light curves. 

\begin{figure}[b]
    \centering
    \includegraphics[width=0.9\linewidth]{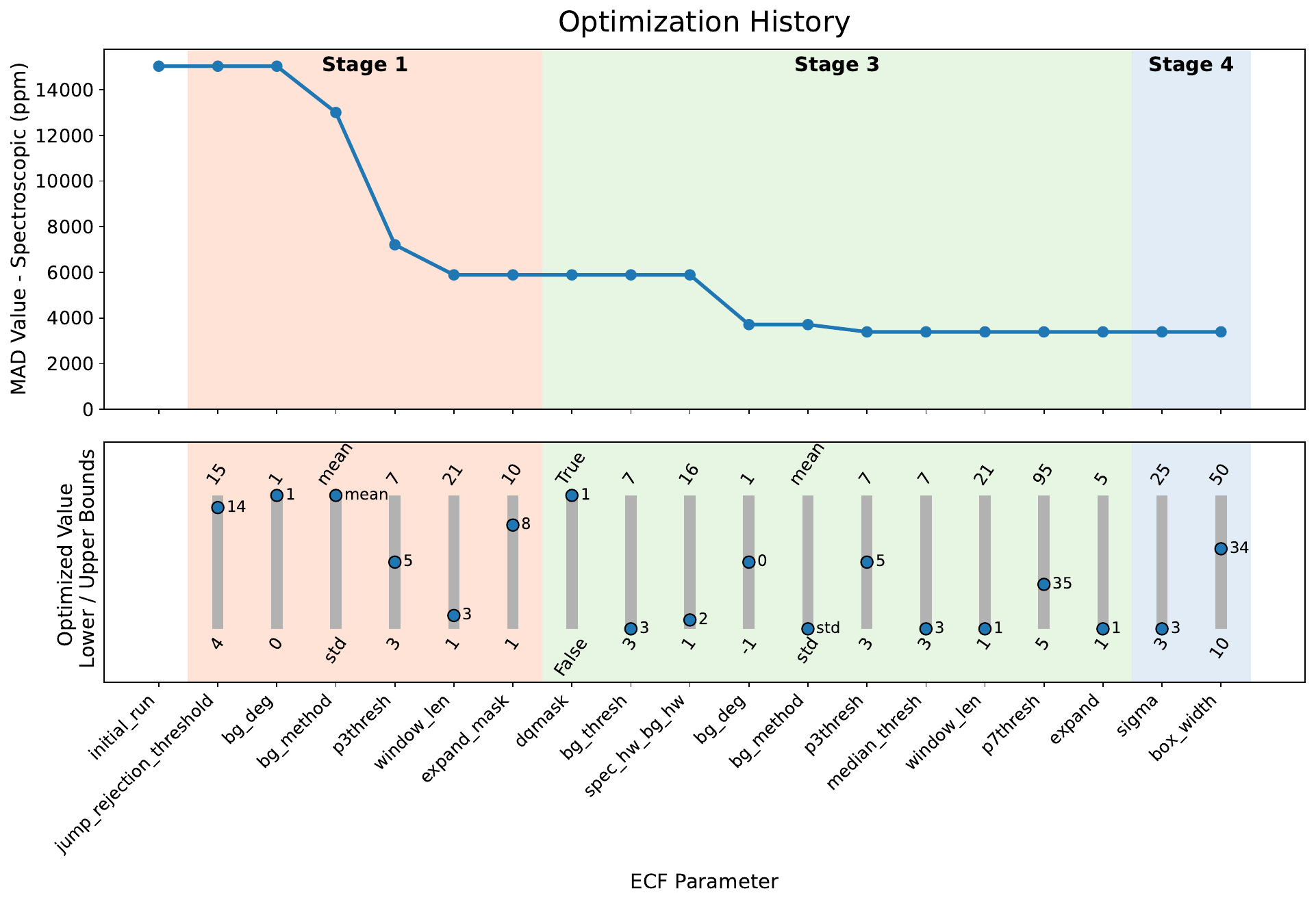}

    \caption{Optimization history for the parameter-by-parameter data reduction with the \textit{optimized} \eureka\ pipeline. \textit{Top:} MAD of the spectroscopic light curves after each successive update to an individual ECF parameter{ForestGreen}{, with the caveat of \texttt{spec\_hw} \& \texttt{bg\_hw} being evaluated in a nested-loop; coincidentally both parameters yielded an optimized value of 2 in this optimization.} Shaded regions indicate the subsets of parameters adjusted in Stages~1, 3, and 4 of the reduction. \textit{Bottom:} Final optimized value for each parameter relative to its allowed bounds, with the lower and upper bounds annotated below and above each slider.
    \vspace{0.2cm}
    \\
    \noindent
    \textit{NOTE -- the \eureka\ labels shown in the x-axis are provided verbatim
    from their ECF parameter namings (e.g. \texttt{spec\_hw}: the half-width of aperture region for spectral extraction). A detailed explanation and units for each parameter is
    available at \cite{eureka-ecf}.}}

    \label{fig:optimization}
\end{figure}

\autoref{fig:optimization} shows the optimization history, where the MAD steadily decreases as successive parameters are tuned, indicating convergence toward a stable reduction. Optimized ECF parameter values and their upper and lower bounds are shown below each MAD value for reference. This plot provides a visual demonstration of how the optimizer identifies parameter combinations that minimize scatter and suppress systematics, ultimately yielding cleaner light curves for downstream fitting. The optimization history shown here highlights the effectiveness of the automated strategy in refining the \textit{optimized} \eureka\ reductions.

\autoref{tab:fit_parameters} lists the best-fit transit and stellar spot parameters derived from our light curve modeling. Both the \textit{optimized} \eureka $+$ \texttt{fleck} and \textit{manual} \eureka $+$ \texttt{spotrod} star spot models are reported, with values shown for three separate visits.

\begin{deluxetable*}{llrrrrrr}
\tabletypesize{\scriptsize}              
\tablewidth{\textwidth}                  
\setlength{\tabcolsep}{3pt}              
\tablecaption{Best-fit transit and star spot model parameters. \label{tab:fit_parameters}}
\tablehead{
Parameter & Unit &
\multicolumn{3}{c}{\textit{Optimized} \texttt{Eureka!\hspace{0.05cm}+\hspace{0.05cm}fleck}} & 
\multicolumn{3}{c}{\textit{Manual} \texttt{Eureka!\hspace{0.05cm}+\hspace{0.05cm}spotrod}} \\
& & Visit 1 & Visit 2 & Visit 3 & Visit 1 & Visit 2 & Visit 3
}
\startdata
R$_p$/R$_s$ & \nodata & 0.153 $\pm$ 8$\times$10$^{-5}$ & 0.154 $\pm$ 8$\times$10$^{-5}$ & 0.153 $\pm$ 2$\times$10$^{-4}$ & 0.153 $\pm$ 2$\times$10$^{-4}$ & 0.153 $\pm$ 6$\times$10$^{-4}$ & 0.153 $\pm$ 2$\times$10$^{-4}$ \\
Period, $P$ & days    & \multicolumn{3}{c}{2.7886556}  & \multicolumn{3}{c}{2.7886550} \\
$t_0-60531|60537|60540$ & MJD & 0.9381115 $\pm$ 8$\times$10$^{-6}$ & 0.5154194 $\pm$ 8$\times$10$^{-6}$ & 0.3040276 $\pm$ 8$\times$10$^{-6}$ & 0.9381133 $\pm$ 8$\times$10$^{-6}$ & 0.5154072 $\pm$ 7$\times$10$^{-6}$ & 0.3040300 $\pm$ 7$\times$10$^{-7}$ \\
Inclination, $i$ & deg & 88.154 $\pm$ 0.013 & 88.131 $\pm$ 0.013 & 88.190 $\pm$ 0.026 & 88.228 $\pm$ 0.030 & 88.196 $\pm$ 0.029 & 88.158 $\pm$ 0.028 \\
$a/R_s$ & \nodata & 12.030 $\pm$ 0.012 & 12.045 $\pm$ 0.011 & 12.033 $\pm$ 0.020 & 12.068 $\pm$ 0.027 & 12.072 $\pm$ 0.023 & 12.028 $\pm$ 0.019 \\
$e$ & \nodata & \multicolumn{3}{c}{0} & \multicolumn{3}{c}{0} \\
$\omega$ & deg & \multicolumn{3}{c}{90} & \multicolumn{3}{c}{90} \\
\hline
$u_1$ & \nodata & 0.125 $\pm$ 0.006 & 0.098 $\pm$ 0.004 & 0.179 $\pm$ 0.012 & 0.167 $\pm$ 0.007 & 0.129 $\pm$ 0.009 & 0.148 $\pm$ 0.014 \\
$u_2$ & \nodata & 0.277 $\pm$ 0.018 & 0.219 $\pm$ 0.019 & 0.229 $\pm$ 0.024 & 0.210 $\pm$ 0.015 & 0.288 $\pm$ 0.027 & 0.274 $\pm$ 0.035 \\
\hline
spot contrast & \nodata & 0.960 $\pm$ 0.008 & 0.971 $\pm$ 0.002 & \nodata & 0.769 $\pm$ 0.032,\; 0.853 $\pm$ 0.207 & 0.609 $\pm$ 0.103 & \nodata \\
spot radius   & \nodata & 0.130 $\pm$ 0.016 & 0.220 $\pm$ 0.010 & \nodata & 0.097 $\pm$ 0.023,\; 0.097 $\pm$ 0.060 & 0.098 $\pm$ 0.081 & \nodata \\
spot latitude & deg & -22.88 $\pm$ 0.05 & -21.20 $\pm$ 0.05 & \nodata & -31.93 $\pm$ 1.68,\; -11.91 $\pm$ 3.98 & -13.51 $\pm$ 1.89 & \nodata \\
spot longitude& deg & -22.80 $\pm$ 0.05 & 36.93 $\pm$ 0.05 & \nodata & -30.12 $\pm$ 1.98,\; 1.51 $\pm$ 1.17 & 30.12 $\pm$ 1.85 & \nodata \\
\enddata
\tablecomments{Orbital and stellar parameters for the primary(\textit{Optimized} \eureka $+$ \texttt{fleck} \citep{Morris2022}), and secondary data reductions(\textit{manual} \eureka $+$ \texttt{spotrod} \citep{Beky2014}). Uncertainties provided are 1$\sigma$ values.}
\end{deluxetable*}

\begin{figure*}[h!]
    \centering
    \includegraphics[width=0.99\linewidth]{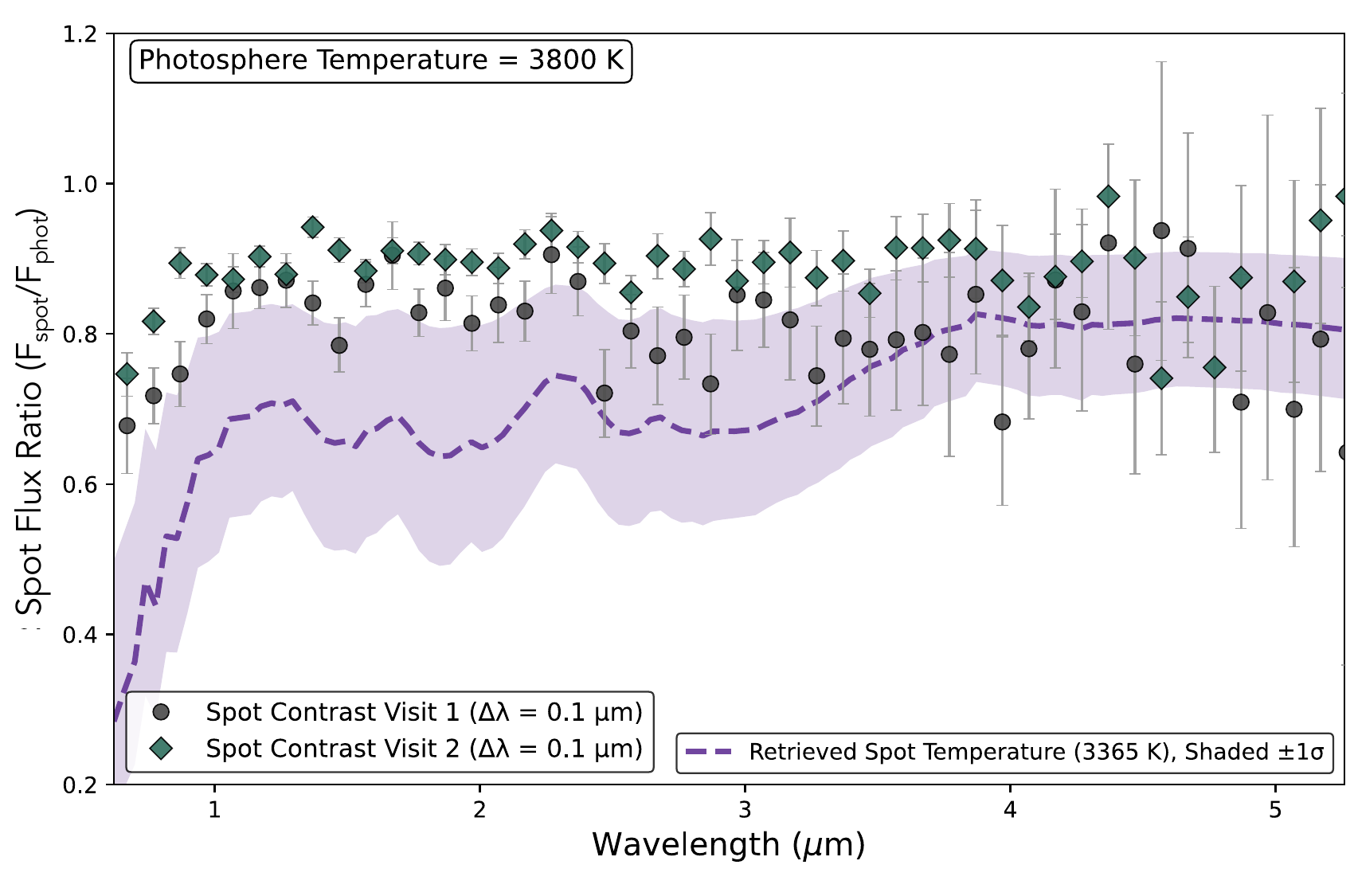}
    \caption{Spot flux ratios assuming a 3800 K stellar photosphere temperature in agreement with both the retrieved photosphere temperature as well as high resolution observations in \citet{Jordan2022}. Spot flux ratios from Visit 1 (\textit{teal}) and Visit 2 (\textit{purple}) derived from the \texttt{spotrod} fits are plotted against wavelength. The expected spot flux ratios derived from the retrieved spot temperature of 3365 K is included for comparison along with the 1$\sigma$ uncertainty region. While the measured spot flux ratios are offset, they still capture the overall structure expected from stellar models.}
    \label{fig:spotcontrasts}
\end{figure*}

\FloatBarrier
\onecolumngrid
Figure~\ref{fig:spotcontrasts} compares the measured spot flux ratios from the manual \texttt{Eureka!} + \texttt{spotrod} star spot models as a function of wavelength across the two spotted visits (Visit 1 and Visit 2). Using \texttt{Phoenix} stellar models, the expected spot flux ratios based on the spot and photosphere temperature are also shown including the 1$\sigma$ uncertainty region. There is a notable 1 to 2$\sigma$ difference in which the spot flux ratios yield a $\sim$200 K hotter spot temperature than the retrieved values of $\sim$3300 K. Given the number of degeneracies in parameter space of the retrievals, this continuum offset in spot temperature is expected. Outside of the offset, the spot flux ratios capture the overall structure expected from stellar models.

\FloatBarrier
\clearpage
\renewcommand\thefigure{\thesection.\arabic{figure}}    
\setcounter{figure}{0}

\section{Additional Retrieval Materials} 
\label{appendix:more_retrievals}

\begin{figure*}[htb!]
    \centering
    \includegraphics[width=0.99\linewidth]{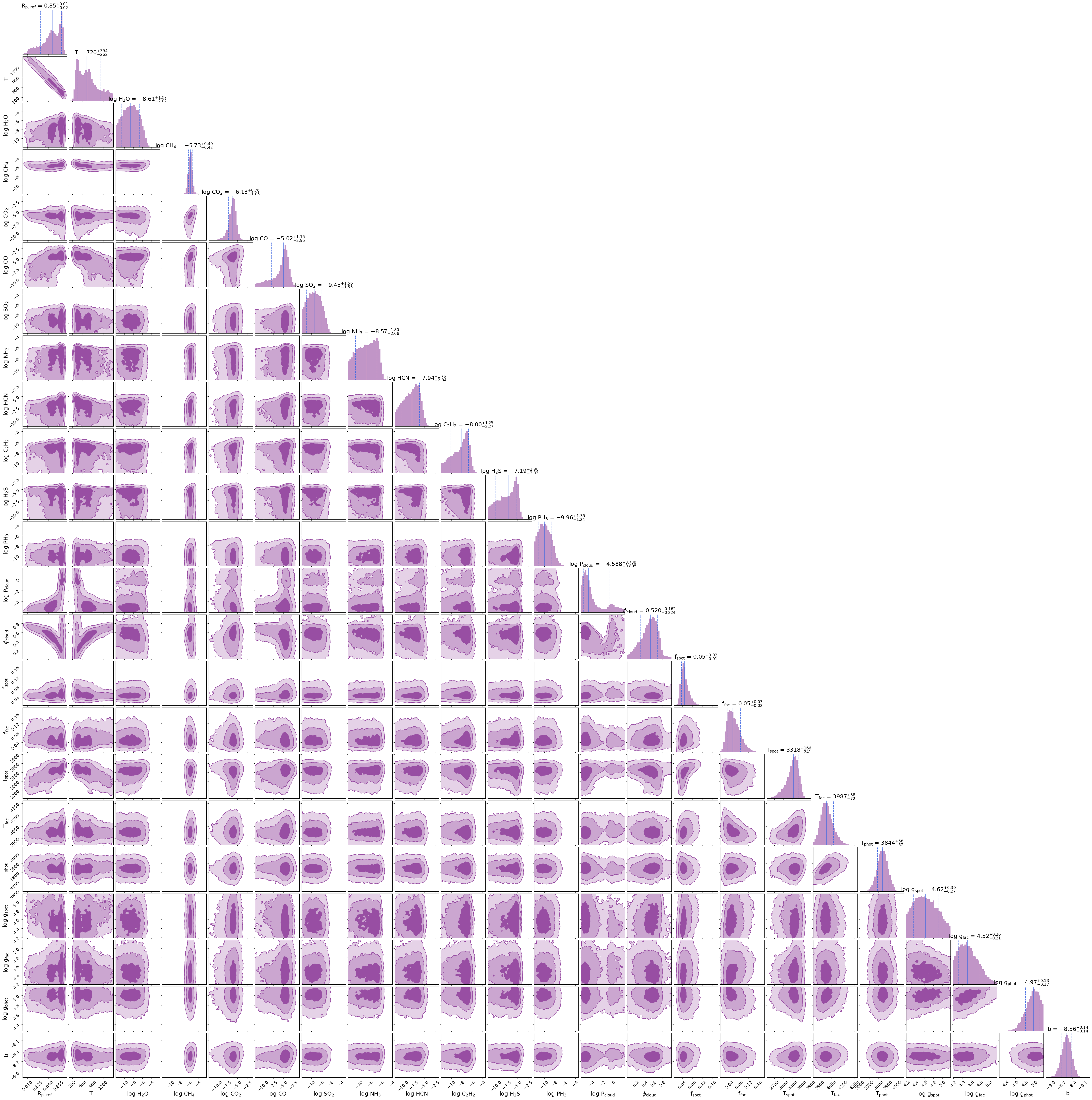}
    \caption{Corner plot showing the \poseidon retrieval posteriors for the Model A (atmosphere + partial clouds + TLS) fit to the coadded transmission spectrum of HATS-75b.}
    \label{fig:corner_modelA}
\end{figure*}

\begin{figure*}
    \centering
    \includegraphics[width=0.99\linewidth]{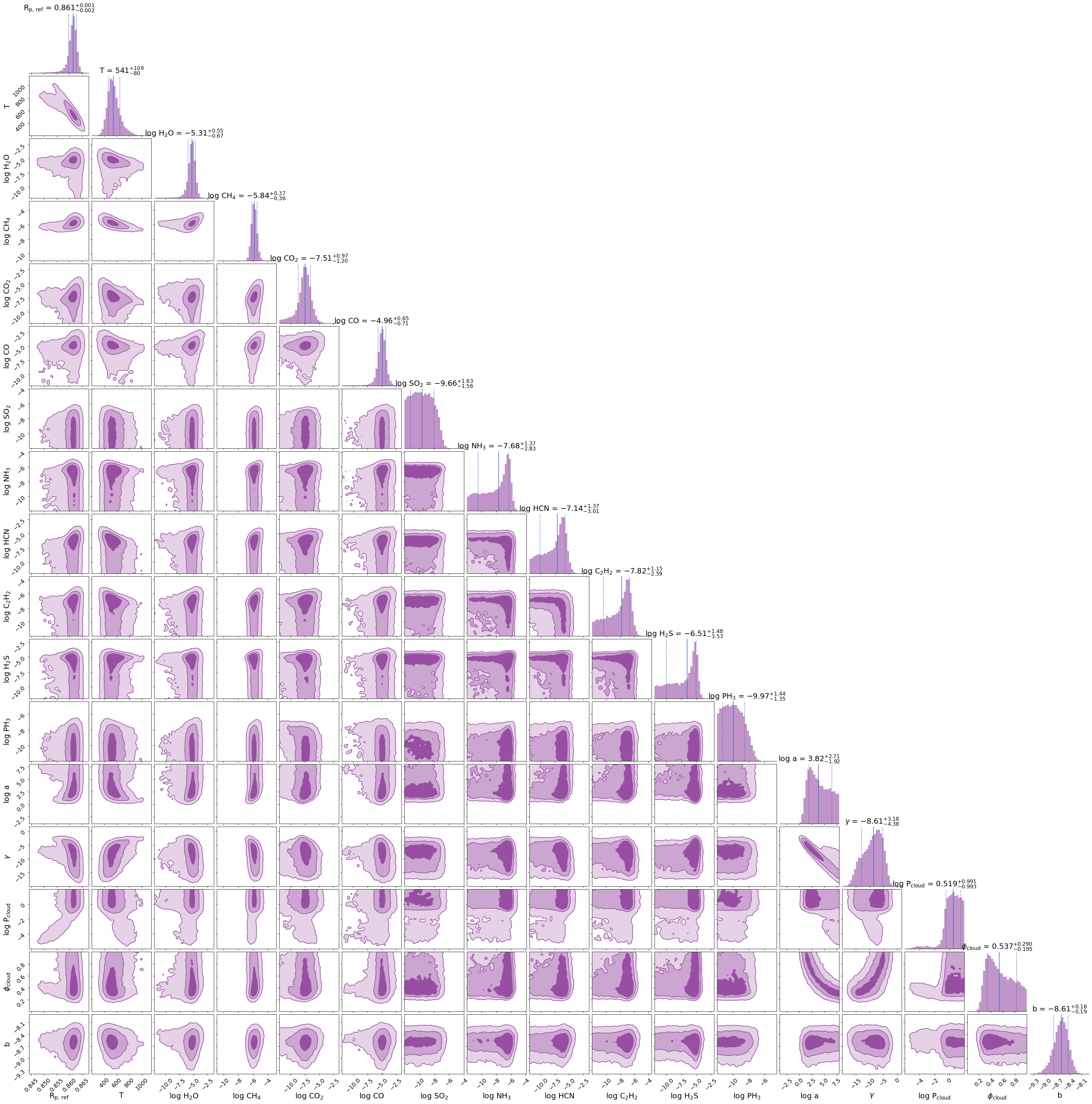}
    \caption{Corner plot showing the \poseidon retrieval posteriors for the Model B (atmosphere + partial clouds + haze) fit to the coadded transmission spectrum of HATS-75b.}
    \label{fig:corner_modelB}
\end{figure*}

This appendix contains supporting material for the retrieval results described in \autoref{sec:retrievals}. \autoref{fig:corner_modelA} and \autoref{fig:corner_modelB} provide the full corner plots for the Model A and Model B retrievals, respectively. 

\clearpage

\bibliography{main, SKLibrary}

\end{document}